\documentclass[useAMS,usenatbib,usegraphicx]{mn2e}
\usepackage[T1]{fontenc}
\usepackage{ae,aecompl}
\usepackage{graphicx}	
\usepackage{amsmath}	
\usepackage{amssymb}	
\usepackage{booktabs,caption,fixltx2e}	
\usepackage{aas_macros} 
\usepackage{upgreek}
\usepackage{url}
\usepackage[usenames]{color}

\newcommand{\kms}{\,km\,s$^{-1}$} 
\newcommand{\Teff}{$T_{\rm eff}$\,}
\newcommand{\logg}{$\log g$\ }

\title[The remarkable star KIC\,11145123]{Spectroscopic and asteroseismic analysis of the remarkable
main-sequence A star KIC\,11145123}
\author[Takada-Hidai et al.]
{Masahide Takada-Hidai$^{1}$\thanks{E-mail: mth\_tsc@tsc.u-tokai.ac.jp}\thanks{Based on data collected at the Subaru Telescope, which is operated by the National Astronomical Observatory of Japan.}
,
Donald W. Kurtz$^{2}$,
Hiromoto Shibahashi$^{3}$
\newauthor{Simon J. Murphy$^{4,5}$, Masao Takata$^{3}$, Hideyuki Saio$^{6}$ and Takashi Sekii$^{7}$}
\\
$^{1}$Liberal Arts Education Center, Tokai University, Kitakaname,
Hirastuka, Kanagawa 259-1292, Japan\\
$^{2}$Jeremiah Horrocks Institute, University of Central
Lancashire, Preston PR1 2HE, UK\\
$^{3}$Department of Astronomy, School of Science, The University of
Tokyo, Bunkyo-ku, Tokyo 113-0033, Japan\\
$^{4}$Sydney Institute for Astronomy, School of Physics, The University
of Sydney, NSW 2006, Australia\\
$^{5}$Stellar Astrophysics Centre, Department of Physics and Astronomy, Aarhus University, 8000 Aarhus C, Denmark\\
$^{6}$Astronomical Institute, Graduate School of Science, Tohoku University, Sendai   980-8578, Japan \\
$^{7}$National Astronomical Observatory of Japan, 2-21-1 Osawa, Mitaka, Tokyo 181-8588, Japan}

\begin{document} 

\maketitle 

\begin{abstract}
A spectroscopic analysis was carried out to clarify the properties of KIC\,11145123 -- the first main-sequence star with a determination of core-to-surface rotation --  based on spectra observed with the High Dispersion Spectrograph (HDS) of the Subaru telescope. The atmospheric parameters ($T_{\rm eff} = 7600$\,K, $\log g = 4.2$, $\xi = 3.1$\,km~s$^{-1}$ and  $ {\rm [Fe/H]} = -0.71$\,dex), the radial and rotation velocities, and elemental abundances were obtained by analysing line strengths and fitting line profiles, which were calculated with a 1D LTE model atmosphere. The main properties of KIC\,11145123 are: (1) A low $ {\rm [Fe/H]} = -0.71\pm0.11$\,dex and a high radial velocity of $-135.4 \pm 0.2$\,km\,s$^{-1}$. These are remarkable among late-A stars. Our best asteroseismic models with this low [Fe/H] have slightly high helium abundance and low masses of 1.4\,M$_\odot$. All of these results strongly suggest that KIC\,11145123 is a Population~II blue straggler; (2) The projected rotation velocity confirms the asteroseismically predicted slow rotation of the star; (3) Comparisons of abundance patterns between KIC\,11145123 and Am, Ap, and blue stragglers show that KIC\,11145123 is neither an Am star nor an Ap star, but has abundances consistent with a blue straggler. We conclude that the remarkably long 100-d rotation period of this star is a consequence of it being a blue straggler, but both pathways for the formation of blue stragglers -- merger and mass loss in a binary system -- pose difficulties for our understanding of the exceedingly slow rotation. In particular, we show that there is no evidence of any secondary companion star, and we put stringent limits on the possible mass of any such purported companion through the phase modulation (PM) technique. 
\end{abstract} 


\section{Introduction} 
\label{sec:intro}

The {\it Kepler} mission provided photometry of unprecedented precision from which the core-to-surface rotation rate of a main-sequence star was measured for the first time in a model-independent manner \citep{Kurtz2014}. The A-type star, KIC\,11145123, pulsates in g\:modes that are sensitive to the outer core, and p\:modes that are  sensitive to the near-surface conditions. Both types of pulsation modes show rotational splitting, from which \citet{Kurtz2014} inferred nearly rigid rotation with a period near 100\,d, and, surprisingly, with the surface rotating slightly, but significantly, faster than the core. \citet{gizon2016} then used the data of \citet{Kurtz2014} to find differences in the splitting in the frequencies of the quadrupole p~modes in this star that led to the discovery that the star is {\it less} oblate than expected, even for such a slow rotator. \citet{gizon2016} speculated that a magnetic field may have suppressed some of the rotational oblateness, and they announced the star as ``the roundest object in the universe''. Since the work of \citet{Kurtz2014}, similar analyses have been carried out for other A/F stars \citep{Saio2015,Schmid2016,Murphy2016}, each showing near-uniform core-to-surface rotation in stars near the terminal age main-sequence (TAMS).

The results of those papers suggest that if  A--F stars typically rotate nearly rigidly at the end of main sequence, a strong angular momentum transport mechanism must operate during the main sequence evolution to produce such rigid rotation, and a mechanism such as internal gravity waves or mass accretion must exist to accelerate the rotation rate of the surface beyond that of the core. 

The rotation period of KIC\,11145123 is exceptionally long for a late-A star, and an explanation is needed for this slow rotation, along with the unexpected rigid rotation.   \citet{Kurtz2014} posited some explanations and ruled out the possibility that the star is a magnetic Ap star because no known Ap star shows both g\:modes and low overtone p\:modes. They also rejected the possibility that the star is in a binary system synchronously rotating with an orbital period of 100\,d, using the frequency modulation (FM) technique \citep{Shibahashi2012}. Instead, they suggested that the star could be either an Am star or a blue straggler, because all Am stars and many blue stragglers are known to rotate slowly \citep[e.g.][]{Lovisi2013, Lovisi2013a, Glaspey1994}.

To investigate their suggestion and clarify the properties of KIC\,11145123, we performed a spectroscopic analysis of this star based on high-dispersion spectra obtained with the Subaru telescope. We also found new asteroseismic models based on our improved knowledge of the star's fundamental parameters and abundances. We find from the abundances and space velocity that the star is Population~II, and our asteroseismic analysis requires some mass loss during the lifetime of the star, hence is consistent with the slow rotation. A remaining puzzle is the mechanism of that mass loss, since we are able to constrain any possible companion to sub-stellar mass, hence the star is not currently in a suitable binary to explain its earlier mass loss. 

\section{Spectroscopic observations and data reduction}
\label{sec:observations}

Our target, KIC\,11145123, was observed with the High Dispersion Spectrograph  \citep[HDS;][]{Noguchi2002} on the Subaru telescope on 2015 July 3. Echelle spectra were obtained in a standard StdYc setup covering the wavelength range of 4390 -- 7120 \AA, and using the image slicer \#3 with $1\times1$ binning, which yields a wavelength resolution of 160\,000. Four spectra were obtained, each with an exposure time of 2000\,s, yielding a total exposure time of 8000\,s. Basic data and the derived atmospheric parameters for our target are shown in Table~\ref{tab:01}, along with the parameters of our best asteroseismic model derived in Section\,\ref{sec:model}.

Standard data reduction procedures (bias subtraction, background subtraction, cosmic ray removal, flat-fielding, extraction of 1D spectra, wavelength calibration, co-addition of spectra, combination into a single spectrum, and normalisation) were carried out with the {\sc iraf} \'echelle package\footnote{{\sc iraf} is distributed by the National Optical Astronomy Observatory, which is operated by the Association of Universities for Research in Astronomy, Inc., under cooperative agreement with the
National Science Foundation.}. The resultant spectra consist of a blue part with wavelength coverage of 4400 -- 5720\,\AA, and a red part with 5800 -- 7120\,\AA.  Signal-to-noise ratios (S/N) were measured on peak continua of each order, and  found for both blue and red parts to be 100 -- 120. To make a  Doppler correction, the radial velocity was measured using 64
 Fe~{\sc i} and 24 Fe~{\sc ii} lines. The radial velocity averaged with weights of line numbers is $V_{\rm r} = - 145.4 \pm 0.2$\,km\,s$^{-1}$, giving a heliocentric radial velocity of  $V_{\rm h} = - 135.4 \pm 0.2$\,km\,s$^{-1}$, as listed in Table~\ref{tab:01}.

\begin{table}
\caption{Fundamental data and model parameters of KIC\,11145123. }
\centering
\begin{tabular}{lr}
 \hline
 Right Ascension (J2000.0)    & $19^{\rm h} 41^{\rm m} 25^{\rm s}$ \\
 Declination  (J2000.0)       & $+48\degr 45\arcmin 15\arcsec$\\
  Kp (mag)${}^{*}$       & 13.123 \\
 \hline
 \multicolumn{2}{c}{spectroscopically derived parameters} \\
 \hline
 \Teff (K)    & 7590 $^{+80}_{-140}$ \\
 \logg (cgs)        & $4.22 \pm 0.13$  \\
 $\xi$ (km\,s$^{-1}$)& $3.1 \pm 0.5$ \\
 $[{\rm Fe/H}] $      & $-0.71 \pm 0.11$  \\
 $V_{\rm h}$ (km~s$^{-1}$)         & $-135.4 \pm 0.2$   \\
 \hline
 \multicolumn{2}{c}{asteroseismically derived parameters} \\
 \hline
$M$ (mass)   & 1.4\,M$_\odot$ \\
 $X$ (H fraction)    & 0.70\\
 $Y$  (He fraction)    & 0.297\\
 $Z$   (metal fraction)   & 0.003 \\
\bottomrule
\end{tabular}
{\footnotesize \\
\vspace{1mm}$^{*}$ Kp is the white light $\it Kepler$ magnitude taken from the revised {\it Kepler} Input Catalog (KIC) \citep{Huber2014}.}
\label{tab:01}
\end{table}

\section{Atmospheric parameters}
\label{sec:atmosphere}

The parameters of the atmospheric model of our target, effective temperature ($T_{\rm eff}$), surface gravity ($\log g$), microturbulence ($\xi$), and metallicity ($[{\rm Fe/H}]),$\footnote{$[{\rm X/Y}] \equiv \log ({\rm X/Y})_{\rm star} - 
\log ({\rm X/Y})_{\sun}$ is used throughout the paper.} were determined by following the principle and algorithm of iteration
procedures described in \citet{Takeda2002}.  Their iteration procedures are based on the  equivalent widths ($W_{\lambda}$) and abundances determined from each of the selected Fe~{\sc i} and Fe~{\sc ii}  lines, and find a final solution for the atmospheric parameters in the ($T_{\rm eff}$, $\log g$, $\xi$) parameter space by adjusting the three parameters from their initial values by small amounts. Each perturbation also results in a new [Fe/H].

In this procedure we selected 67 Fe~{\sc i} and 15 Fe~{\sc ii} lines from \citet{Westin2000}, and also 6 Fe~{\sc ii} lines from \citet{Takeda2005}. The $gf$-values of these Fe lines were also adopted from these sources. To analyse the Fe lines, we used the {\sc sptool} software package developed by Y. Takeda\footnote{\url{http://optik2.mtk.nao.ac.jp/~takeda/sptool/}}, which is based on Kurucz's {\small ATLAS9} 1D LTE model atmospheres and {\small WIDTH9} program for abundance analysis \citep{Kurucz1993}. The program {\sc spshow} was used to measure $W_{\lambda}$ by Gaussian profile fitting in
most cases, but by direct integration in others, and the program {\sc width} was used to derive abundances from the $W_{\lambda}$ of selected Fe lines. The selected lines of Fe~{\sc i} and Fe~{\sc ii} are given in Appendix\,\ref{app:fe} in Tables\:\ref{tab:02} and \ref{tab:03} along with the atomic constants and measured equivalent widths. 

We obtained our solution for the atmospheric parameters using a standard iterative method as follows. We began with atmospheric parameters for our starting model selected from the {\it Kepler} Input Catalogue (KIC) as revised by \citet{Huber2014}. At each iteration we initially kept \Teff and \logg fixed, and determined $\xi$ by insisting that the abundances derived from the  Fe~{\sc i} lines were independent of the corresponding $W_{\lambda}$. This resulted in a new [Fe/H] that was used to determine \Teff under the requirement of excitation equilibrium, which specifies that the Fe\,{\sc i} abundances are independent of the lower excitation potential, $\chi$. The final step in the iteration was to evaluate $\log g$ under the requirement of ionization equilibrium, such that the Fe abundance derived from Fe\,{\sc i} and Fe\,{\sc ii} lines are equal. The next iteration used the parameters thus obtained as the new starting model, until convergence was reached. The best-fitting parameters are: $T_{\rm eff} = 7590^{+80}_{-140}$\,K, \logg= 4.22 $\pm 0.13$, $\xi = 3.1 \pm 0.5$, and $[{\rm Fe/H}] = -0.71\pm0.11$.

As a check of these parameters, we calculated the reduced chi-squared statistic, $\chi^2_{\rm r}$, for 16 models created from the four parameters perturbed by their uncertainties. We adopted the set of observed $W_{\lambda}$ of 67 Fe~{\sc i} and 21 Fe~{\sc ii} lines from Tables\:\ref{tab:02} and \ref{tab:03}, and calculated $W_{\lambda}$ of each line for each model with 88 degrees of freedom. None of those 16 models had $\chi^2_{\rm r}$ lower than that of the best-fitting model, for which we found $\chi^2_{\rm r} = 3.20$.

\section{Rotation velocity}
\label{sec:rotation}

The exceptionally long rotation period near 100\,d of KIC\,11145123 derived asteroseismically by  \citet{Kurtz2014} 
to predict a surface equatorial rotation velocity of $v_{\rm eq} = 1$\,km\,s$^{-1}$, implying $v \sin i \le 1$\,km\,s$^{-1}$. However, observational measures of $v \sin i$ are made from line broadening, which includes macroturbulent broadening, as well as pulsational broadening from both the p~modes and the g~modes in this star. These additional broadening mechanisms are discussed,  modelled and evaluated by \citet{Murphy2016} in the context of another relatively slowly rotating $\gamma$~Dor star, KIC\,7661054, where they find that the combination of the pulsational and macroturbulent broadening are of the order of several km\,s$^{-1}$, and certainly less than a total of 10\,km\,s$^{-1}$. It is common usage to give the line broadening measurement as if it were only $v \sin i$, and this is a good approximation for moderate to fast rotating stars. However, when the equatorial rotation velocity is of the order of a few km~s$^{-1}$, then it is necessary to be aware that the line broadening originates from these multiple sources. 

A-type stars with normal abundances in general rotate relatively fast, with $v \sin i > 100$\,km\,s$^{-1}$. Even the chemically peculiar Ap and Am stars, which generally have $v \sin i < 100$\,km\,s$^{-1}$, still often show equatorial rotation velocities of tens of km\,s$^{-1}$. Thus the asteroseismic prediction that the total line broadening from rotation, pulsation and macroturbulence of $v \sin i \le 10$\,km\,s$^{-1}$ for KIC\,11145123 is a strong prediction that is easily falsified, should it be incorrect -- in particular, should the rotation period be significantly less than the 100\,d measured asteroseismically by \citet{Kurtz2014}. Therefore to test for the predicted slow rotation of our target, we performed line profile fitting to derive the convolved line broadening velocities.

The line profile fittings were made based on the final model atmosphere, using the program {\sc mpfit} in the {\sc sptool} software package under the following assumptions as described by \citet{Takeda2008b}:

(1) The observed spectrum is a convolution of the modelled intrinsic spectrum and the total macrobroadening function, $f_{\rm T}$.

(2) $f_{\rm T}$ is a convolution of three functions: the instrumental broadening $f_{\rm i}$; the projected rotation broadening, $f_{\rm r}$; and the macroturbulence broadening (which includes the pulsational broadening), $f_{\rm m}$. Thus, $f_{\rm T} =  f_{\rm i} * f_{\rm r} * f_{\rm m}$ (* denotes convolution). 

(3) We approximate these broadening functions with the same Gaussian form, parametrized by the $e$-folding half-width ($v_{k}$) as \mbox{$f_{k}\propto \exp(-v^{2}/v_{k}^{2})$}, where $k$ represents any of the suffixes. These broadening parameters are thus related as $ v_{\rm T}^{2} \equiv v_{\rm i}^{2} + v_{\rm r}^{2} + v_{\rm m}^{2}$.

(4) The combined broadening function, $f_{\rm rm}$, of the projected rotation broadening and macroturbulence broadening functions is defined as $f_{\rm rm} \equiv f_{\rm r}*f_{\rm m}$, together with the relation of the combined broadening parameter, $v_{\rm rm} \equiv \sqrt{v_{\rm T}^{2} - v_{\rm i}^{2}} $ $= \sqrt{ v_{\rm r}^{2} + v_{\rm m}^{2}}$.

The instrumental broadening was estimated to be $v_{\rm i}$ $=$ $c_{0}/(2\sqrt{\rm \ln 2}\,R) = 1.13 $\,km\,s$^{-1}$, where $c_{0}$ is the speed of light and $R$ the spectral resolution (160\,000), which is derived from measurements of the full width at half maximum of emission lines of the Th-Ar comparison spectrum. 

\subsection{Confirmation of an argument by \citet{Landstreet2009}}
\label{sec:confirmation}

To confirm the validity of our fitting procedure to obtain $v_{\rm T}$ and $v_{\rm rm}$, we first made a test calculation for the
lines studied by \citet{Landstreet2009} to test in KIC\,11145123 their argument that for stars with \Teff less than about 10\,000\,K, $v \sin{i}$ is dependent on line strengths, $W_\lambda$, so that the signatures of local velocity fields may become more evident in stronger lines.

They used 10 selected lines of Cr~{\sc ii}, Ti~{\sc ii}, Fe~{\sc i}, and Fe~{\sc ii} to investigate the atmospheric velocity fields in a sample of sharp-lined main-sequence A stars. They found 8 main-sequence A stars with \Teff $< 10\,000$ K in which 
evidence of velocity fields is directly detected in line profiles showing depressed blue wings and a discrepancy between observed and theoretical line shapes (see their Table 2). Among these A stars, $v \sin{i}$ values were also estimated; they found that the measured $v \sin{i}$ is smaller when derived from weak lines than from strong lines, indicating a dependence of $v \sin{i}$ on line strengths. They speculated that velocity fields are probably convective motions reaching the atmosphere.

To confirm the dependence of $v \sin{i}$ on line strengths, we adopted  four lines (Cr~{\sc ii} 4554.99\,\AA\ and 4634.07\,\AA, Ti~{\sc ii} 4563.76 \AA, and Fe~{\sc i} 4637.50 \AA) used by \citet{Landstreet2009}, and measured $W_{\lambda}$ of these lines in the observed  spectrum of our target. The values of  $v_{\rm T}$ and $v_{\rm rm}$ were derived from profile fittings to these lines, based on the atmospheric model of our target. The $gf$\ values of these lines were adopted from \citet{Castelli2004}. The uncertainties of  $v_{\rm rm}$  were estimated from the simulations of profile fittings to each line for several values of  $v_{\rm T}$.

\begin{table}
\centering
\caption{
The measured values of  $v_{\rm T}$ and $v_{\rm rm}$ for four lines derived from the observed spectrum of our target, KIC\,11145123. The lines are listed in order of increasing $W_\lambda$, making the correlation between $W_\lambda$ and $v_{\rm rm}$ obvious. } 
\begin{tabular}{rrrrr}
 \toprule
\multicolumn{1}{c}{Line}  &  \multicolumn{1}{c}{$W_{\lambda}$} & \multicolumn{1}{c}{$v_{\rm T}$}  &  \multicolumn{1}{c}{$v_{\rm rm}$}    \\
      &   \multicolumn{1}{c}{m\AA}       &  \multicolumn{1}{c}{\kms}  &  \multicolumn{1}{c}{\kms}          \\
\hline
Fe~{\sc i}  4637  &  17.8  &  6.00  & $5.89  \pm 1.15 $ \\
Cr~{\sc ii} 4554  &  33.0  &  6.69  & $6.59   \pm 0.43 $ \\
Cr~{\sc ii} 4634  &  42.0  &  6.75  & $6.66   \pm 0.33 $ \\
Ti~{\sc ii} 4563  & 174.2  &  7.54  & $7.45   \pm 0.07 $ \\
\bottomrule
\end{tabular}
\label{tab:broadening}
\end{table}

The results are shown in Table\,\ref{tab:broadening}, where the values of $v_{\rm rm}$ are noticeably dependent on $W_{\lambda}$, indicating that $v \sin{i}$ depends on line strengths as argued by \citet{Landstreet2009}. Because our target has $T_{\rm eff} = 7590 ^{+80}_{-140}$\,K, the comparison with the sample in \citet{Landstreet2009} is valid. Consequently, we verified our fitting procedure and arrived at the conclusion that $v_{\rm rm}$, corresponding to $v \sin{i}$, is least influenced by local velocity fields in the weak line limit as $W_{\lambda} \rightarrow 0 $ m\AA.

\subsection{Investigating line broadening and the rotational velocity}
\label{sec:investigating}

\begin{figure}
\centering
\includegraphics[width=0.9\columnwidth]{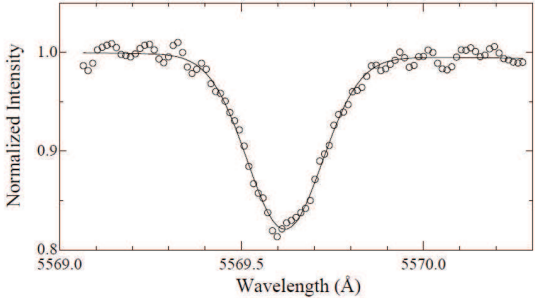}
\caption{An example of the typical profile fitting for the Fe~{\sc i} 5569\,\AA\ line with $W_{\lambda}$ $= 44.4$\,m\AA. Open circles are the observations; the solid line shows the theoretical profile converged after five iterations. The converged solution gives $v_{\rm T} = 6.7$\,km\,s$^{-1}$ and $v_{\rm rm} = 6.6$\,km\,s$^{-1}$, and also the abundance of $\log$ Fe~{\sc i} = 6.73.}
\label{Fig:profile}
\end{figure}

To disentangle the line-broadening contributions, we selected 51 lines from the 67 lines used in Section\,\ref{sec:atmosphere} by adopting those with the least-blended (most symmetric) profiles. The profile fitting calculations yielded the results of $v_{\rm T}$ and then $v_{\rm rm}$. An example of the typical fitting is shown for the Fe~{\sc i} 5569\,\AA\ line with its $W_{\lambda} = 44.4$\,m\AA\ in Fig.\,\ref{Fig:profile}. The converged solution of fitting gives $v_{\rm rm} = 6.6$\,km\,s$^{-1}$\, and an abundance of log Fe~{\sc i} = 6.73, which agrees  well with the final abundance of log Fe = 6.79 (corresponding to [Fe/H] = $-0.71$), demonstrating that our fitting procedure works well.

\begin{figure}
\centering
\includegraphics[width=0.9\columnwidth]{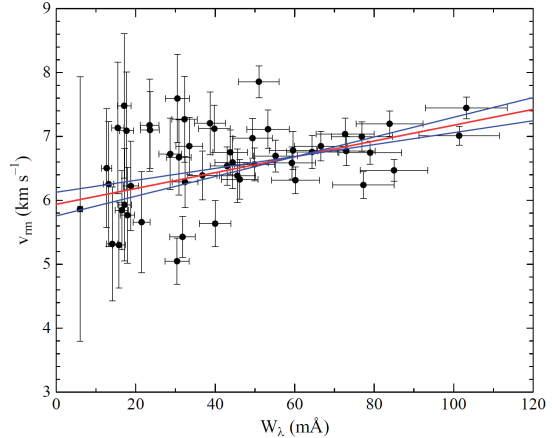}
\caption{The values of $v_{\rm rm}$ derived from 51 Fe~{\sc i} lines are plotted as a function of  $W_{\lambda}$ together with the error bars of both $W_{\lambda}$ (assuming a 10~percent error) and $v_{\rm rm}$.  A least-squares fit is derived taking these error bars into account. The best fit is depicted in a red line with the intercept of 5.94\,km~s$^{-1}$, and the fits corresponding to $\pm 1{\sigma}$ errors are shown in the blue lines with the intercepts of 5.76 and 6.13\,km~s$^{-1}$, respectively. }
\label{Fig:linear}
\end{figure}

The behaviour of $v_{\rm rm}$ derived from 51 lines is shown as a function of $W_{\lambda}$ in Fig.\,\ref{Fig:linear}.
When  we estimate the rotation velocity $v_{\rm r}$  from the projected rotation velocity $v_{\rm r} \sin{i}$, we must separate the macroturbulence $v_{\rm m}$ from   $v_{\rm rm}$. The separation procedure requires a sophisticated approach such
as the one adopted by \citet{Gray2014}, who analysed $v \sin{i}$ of five early-A slow rotators by deriving the radial-tangential macroturbulence ($\zeta_{\rm RT}$) from the Fourier transforms of the observed line profiles, based on spectra with very high S/N of 370 -- 1840.  His sample stars had  $v \sin{i} = 6.0$ -- 26.2\,km\,s$^{-1}$, along with $\zeta_{\rm RT} = 5.7$ -- 2.0\,km\,s$^{-1}$. For example, $o$ Peg had $v \sin{i} = 6.0$\,km\,s$^{-1}$ and $\zeta_{\rm RT} = 5.7$\,km\,s$^{-1}$; $\alpha$ Dra had $v \sin{i} = 26.2$\,km\,s$^{-1}$ and $\zeta_{\rm RT} = 2.0$\,km\,s$^{-1}$. From these results, \citeauthor{Gray2014} suggested that even with the relatively large uncertainties in $\zeta_{\rm RT}$, there may be a connection between $\zeta_{\rm RT}$ and $v \sin{i}$, in that larger rotation may result in smaller macroturbulence.  However, with the radial-tangential macroturbulence formulation having being set up with solar granulation in mind, Gray was cautious of its application to A stars without large convective envelopes. Taking into account his discussions \citep{Gray2014} and the lower quality (S/N = 100 -- 120) of our spectrum, we did not attempt to separate $v_{\rm m}$ from $v_{\rm rm}$. Since $v_{\rm rm} < 8$\,km\,s$^{-1}$ (Fig.\,\ref{Fig:linear}), $\zeta_{\rm RT}$ may account for the majority of $v_{\rm rm}$.

Following the suggestion described in Section \ref{sec:confirmation} that $v_{\rm rm}$ is least influenced by local velocity fields in the weak-line limit as $W_{\lambda} \rightarrow 0 $ m\AA, we adopted the intercept of the $v_{\rm rm}$ axis of the linear least-squares fit shown in Fig.\,\ref{Fig:linear}, $v_{{\rm rm},0} = 5.9 \pm 0.2\,{\rm km}\,{\rm s}^{-1}$, as the {\it apparent} projected rotation velocity, $v_{\rm a}\sin{i}$. The error bar is the error in the value of the intercept taking into account errors in both coordinates.

It is found that $v_{\rm rm,0}$ $=5.9$\,km\,s$^{-1}$ is consistent with $v_{\rm rm}$ shown in Table\,\ref{tab:broadening}  for the lines of Cr~{\sc ii}, Ti~{\sc ii}, and Fe~{\sc i}.  We also calculated $v_{\rm rm}$ for 19 lines of Fe~{\sc ii}, and found the intercept value of the least-squares fit to the plot of $v_{\rm rm}$ vs $W_{\lambda}$ to be $7.0 \pm 0.3$\,km~s$^{-1}$, which is in acceptable agreement with $v_{\rm rm,0} = 5.9 \pm 0.2$\,km~s$^{-1}$. Hence we are confident of the validity of choosing Fe~{\sc i} lines for the determination of the rotation velocity.
  
Our {\it apparent} projected rotation velocity $v_{\rm a}\sin{i} = 5.9 \pm 0.2$\,km\,s$^{-1}$ confirms that our target 
is a very slow rotator, as \citet{Kurtz2014} found asteroseismically. The $v_{\rm r} \sin{i} = 1$\,km\,s$^{-1}$ predicted by \citet{Kurtz2014} is well-supported. We therefore suggest that the difference between our measured $v_{\rm a}\sin{i}$ and their prediction is significant for exploring macroturbulence and the pulsational velocity fields in the atmosphere of this star. 

\section{Abundance Analysis}
\label{sec:abundances}

Abundances for the selected elements were determined using the {\sc sptool} software package based on the final model atmosphere. Its {\sc spshow} program was used to measure $W_{\lambda}$ of lines of the elements by  Gaussian profile fitting in most cases, but direct integration in others, and the program {\sc width} was used to derive abundances from  $W_{\lambda}$. Atomic data of wavelengths, $\chi$, $\log gf$ values are based on \citet{Kurucz1995}, but when updated $\log gf$ values were available from \citet{Castelli2004}, those were adopted instead. Table \ref{tab:A} in the Appendix shows the atomic data, $W_{\lambda}$, and the abundances derived from each line together with the average abundances for each ion and the number of lines used.

We searched for a magnetic field by applying the method of \citet{Mathys1990}, which uses the ratio of the strengths of the Fe~{\sc ii} 6147.7\,\AA\ and 6149.2\,\AA\ lines to estimate the mean magnetic field modulus $\langle H \rangle$. Within the errors, we found no significant difference in strength between these two magnetically sensitive Fe~{\sc ii} lines, hence our analysis proceeded with the assumption of no magnetic broadening. It should be noted, however, that within the errors we cannot rule out a magnetic field strength of the order of $1$\,kG.

We treated the Li doublet as a single line because only an upper limit for equivalent width could be measured, as noted in Table \ref{tab:A}. The upper limit of $W_{\lambda}$ was estimated to be 0.4 m\AA\, by the \citet{Cayrel1988} formula for 
$\delta x=0.014$ \AA, $w=6708$\,\AA$/R$ (at spectroscopic resolution $R=160\,000$), and S/N$ =110$. 

The abundances of other doublet or triplet lines were obtained using the program {\sc mpfit} via Gaussian profile fitting (denoted ``doublet fit'' or ``triplet fit'' in Table \ref{tab:A}.) When hyperfine splitting (hfs) components and relative isotopic fractions of odd nuclei are both available and significant for a given line, the Gaussian profile fitting takes the hfs components and relative isotopic fractions into account. Test calculations demonstrated that lines with relative isotopic fractions less than 0.1, or with $W_{\lambda}$ $\le 20$ m\AA, yield negligible differences that are less than 0.02\,dex between abundances
with and without hfs components and isotopic fractions. We adopted hfs components and isotopic fractions from \citet{Kurucz2011}\footnote{http://kurucz.harvard.edu/linelists/gfhyperall/} for Sc~{\sc ii} and Mn~{\sc i}, and from \citet{McWilliam1998} for Ba~{\sc ii}. The corresponding lines are labelled as ``hfs fit'' in Table \ref{tab:A}. The resulting abundances of each element are summarised in Table\,\ref{tab:abundances}.

\begin{table*}
\caption{Summary of abundances of each element. Average abundances of log~X for element X are  given in the 3rd column.  Standard deviations of each log~X,  uncertainties due to error bars of \Teff, \logg, and $\xi$ (labelled tg$\xi$ in column 7),
 and total uncertainties are given in the 6th, 7th, and 8th columns, respectively. The numbers of doublet or triplet lines, and lines of hyperfine structure (hfs)  fit used to derive the abundances are given in the remarks column. The uncertain abundances are noted as ``uncertain'' in the remarks. Solar abundances from \citet{Asplund2009} are listed in the last
 column. }
\begin{center}
\begin{tabular}{rlrrrcrrrcr}
 \toprule

 \multicolumn{1}{c}{Code}    &      \multicolumn{1}{c}{Ion}      &     \multicolumn{1}{c}{log X}     &      \multicolumn{1}{c}{[X/H]}      &
 \multicolumn{1}{c}{[X/Fe]}      &     \multicolumn{1}{c}{Std.Dev.}     &      \multicolumn{1}{c}{Error}       &       \multicolumn{1}{c}{Total}       &
 \multicolumn{1}{c}{No. of}    &        \multicolumn{1}{c}{Remark}       &    \multicolumn{1}{c}{Sun} \\
           &               &    average    &                 &
 &     \multicolumn{1}{c}{1$\sigma$}     &       \multicolumn{1}{c}{tg$\xi$}     &       \multicolumn{1}{c}{error}       &   \multicolumn{1}{c}{lines}
 &                     &  \\
  \hline
                         &               &               &           &
 &                 &                  &           &               &
 &     \\
    3.00   &       Li~{\sc i}      &     $\le$ 1.38    &    $\le$  0.33
 &     $\le$ 1.04    &       $\cdots$       &      $\cdots$        &    $\cdots$    &
 1      &      1 doublet      &    1.05  \\
    6.00   &      C~{\sc i}     &      7.82     &     $-$0.61       &      0.10
 &       0.15      &       0.05       &        0.16       &     10
 &      2 doublets     &    8.43  \\
    8.00   &      O~{\sc i}      &      8.38     &     $-$0.31       &      0.40
 &       0.09      &       0.09       &        0.13       &      3
 &      3 triplets     &    8.69 \\
   11.00   &      Na~{\sc i}    &      5.75     &     $-$0.49       &      0.22
 &       0.04      &       0.04       &        0.06       &      4
 &      1 doublet      &    6.24 \\
   12.00   &      Mg~{\sc i}    &      7.13     &     $-$0.47       &      0.24
 &       0.04      &       0.06       &        0.07       &      3
 &                     &    7.60  \\
   12.01   &     Mg~{\sc ii}     &      7.29     &     $-$0.31       &      0.40
 &       $\cdots$       &       0.11       &        0.11       &      1
 &      1 triplet      &    7.60  \\
   14.00   &      Si~{\sc i}    &      7.05     &     $-$0.46       &      0.25
 &       0.22      &       0.04       &        0.22       &     14
 &                     &    7.51  \\
   14.01   &     Si~{\sc ii}    &      7.03     &     $-$0.48       &      0.23
 &       0.07      &       0.08       &        0.11       &      3
 &                     &    7.51  \\
   16.00   &      S~{\sc i}     &      6.87     &     $-$0.25       &   0.46       &       0.08      &       0.03       &   0.09       &      8
 &   1 doublet, 3 triplets  &    7.12  \\
   20.00   &      Ca~{\sc i}    &      5.95     &     $-$0.39       &      0.32
 &       0.12      &       0.07       &        0.14       &     25
 &                     &    6.34  \\
   20.01   &     Ca~{\sc ii}    &      6.02     &     $-$0.32       &      0.39
 &       0.09      &       0.06       &        0.11       &      3
 &                     &    6.34  \\
   21.01   &     Sc~{\sc ii}    &      2.80     &     $-$0.35       &      0.36
 &       0.15      &       0.06       &        0.16       &      9
 &      4 hfs fit      &    3.15 \\
   22.00   &      Ti~{\sc i}    &      4.59     &     $-$0.36       &      0.35
 &       0.11      &       0.06       &        0.13       &     15
 &                     &    4.95  \\
   22.01   &     Ti~{\sc ii}    &      4.59     &     $-$0.36       &      0.35
 &       0.18      &       0.06       &        0.19       &     32
 &                     &    4.95  \\
   24.00   &      Cr~{\sc i}    &      4.90     &     $-$0.74       &     $-$0.03
 &       0.11      &       0.07       &        0.13       &     14
 &                     &    5.64  \\
   24.01   &     Cr~{\sc ii}    &      4.96     &     $-$0.68       &      0.03
 &       0.12      &       0.05       &        0.13       &     21
 &                     &    5.64  \\
   25.00   &      Mn~{\sc i}    &      4.57     &     $-$0.86       &     $-$0.15
 &       0.05      &       0.06       &        0.08       &      7
 &      2 hfs fit      &    5.43  \\
   26.00   &      Fe~{\sc i}    &      6.79     &     $-$0.71       &      0.00
 &       0.10      &       0.08       &        0.13       &     67
 &           &    7.50  \\
   26.01   &     Fe~{\sc ii}    &      6.79     &     $-$0.71       &      0.00
 &       0.14      &       0.08       &        0.16       &     21
 &            &    7.50  \\
   28.00   &      Ni~{\sc i}   &      5.70     &     $-$0.52       &      0.19
 &       0.10      &       0.06       &        0.12       &     41
 &                     &    6.22  \\
   29.00   &      Cu~{\sc i}    &      3.56     &     $-$0.63       &      0.08
 &       0.05      &       0.07       &        0.09       &      2
 &      uncertain      &    4.19  \\
   30.00   &      Zn~{\sc i}    &      4.15     &     $-$0.41       &      0.30
 &       0.03      &       0.06       &        0.07       &      3
 &                     &    4.56  \\
   39.01   &      Y~{\sc ii}    &      1.87     &     $-$0.34       &      0.37
 &       0.06      &       0.07       &        0.09       &      7
 &                     &    2.21  \\
   40.01   &     Zr~{\sc ii}   &      2.38     &     $-$0.20       &      0.51
 &       $\cdots$        &       0.06       &        0.06       &      1
 &      uncertain      &    2.58  \\
   56.01   &     Ba~{\sc ii}   &      1.58     &     $-$0.60       &      0.11
 &       0.04      &       0.15       &        0.15       &      4
 &     4 hfs fit    &    2.18  \\
   58.01   &     Ce~{\sc ii}    &      1.26     &     $-$0.32       &      0.39
 &        $\cdots$       &       0.08       &        0.08       &      1
 &      uncertain      &    1.58  \\
   60.01   &     Nd~{\sc ii}    &      0.91     &     $-$0.51       &      0.21
 &        $\cdots$   &       0.09       &        0.09       &      1
 &      uncertain      &    1.41  \\

\bottomrule
\end{tabular}
\label{tab:abundances}
\end{center}
\end{table*}

\begin{figure}
\centering
\includegraphics[width=0.95\columnwidth]{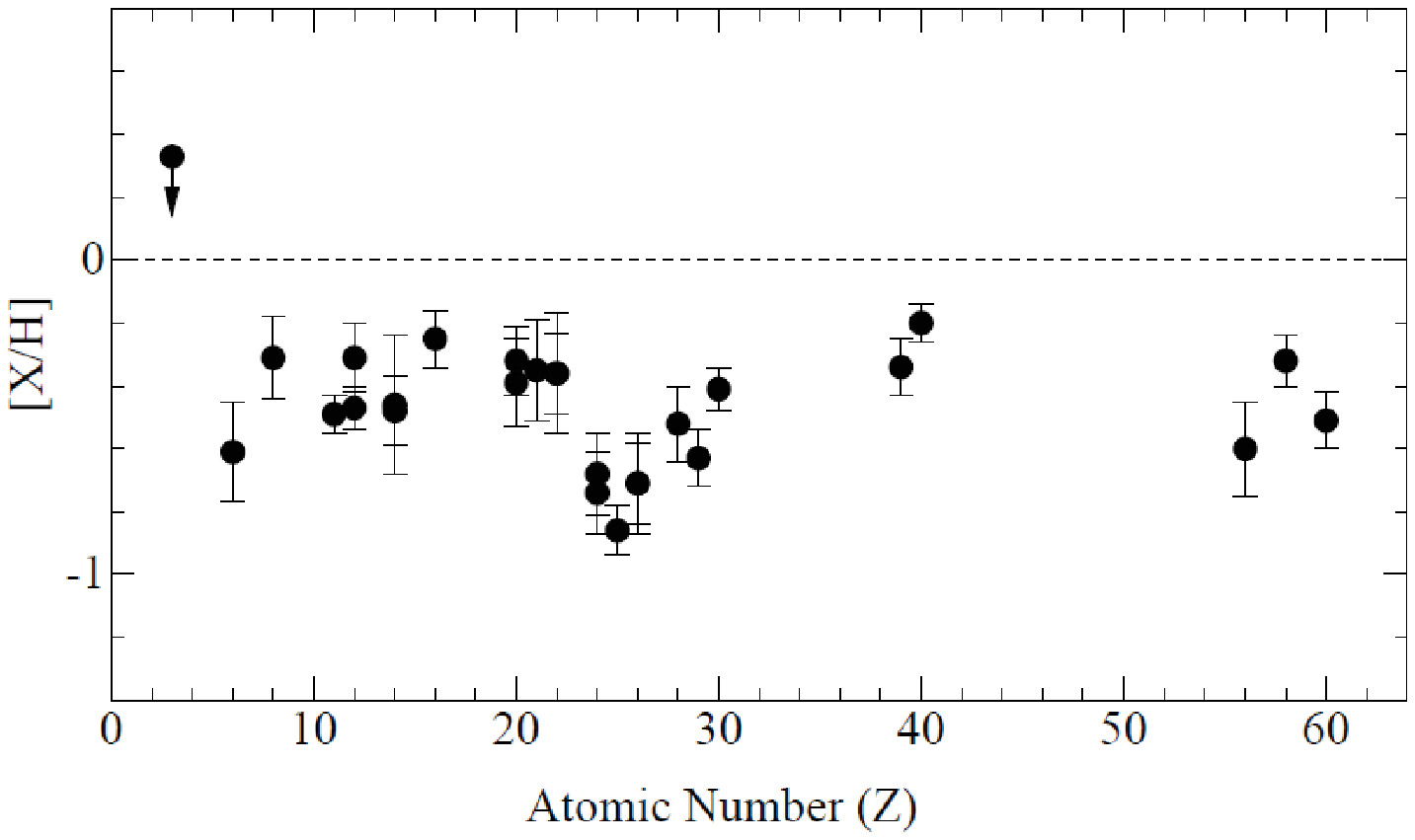}
\includegraphics[width=0.95\columnwidth]{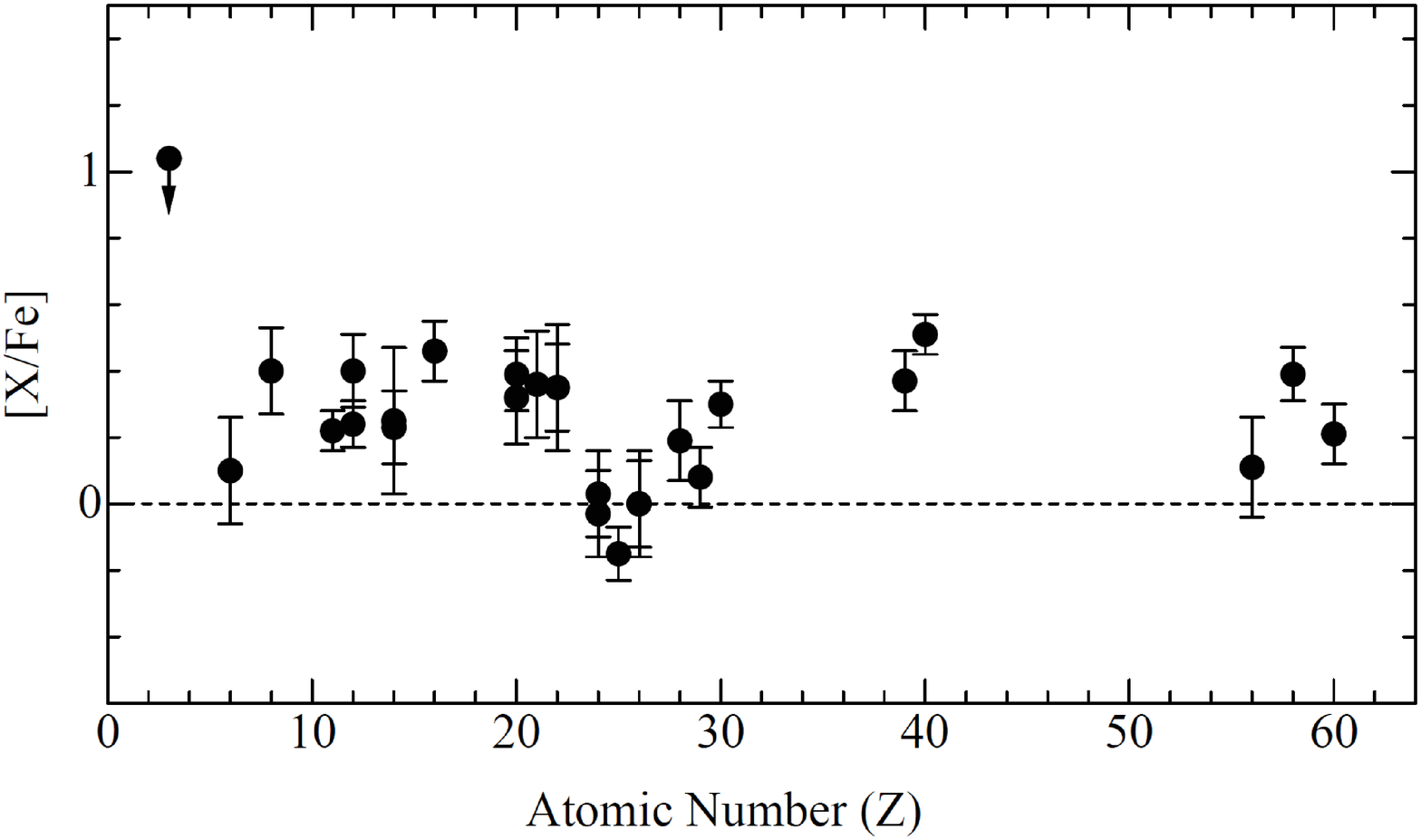}
\caption{Top panel: [X/H] against atomic number for KIC\,11145123. Except for Li, which is derived as an upper limit only, the systematic underabundances are consistent with the remarkably deficient Fe abundance, [Fe/H]$=-0.71$\,dex. Bottom panel: [X/Fe] against atomic number  for KIC\,11145123.}
\label{Fig:abundances}
\end{figure}

To illustrate the abundance pattern relative to the Sun, we depicted [X/H] against atomic number in the top panel of Fig.\,\ref{Fig:abundances}. All elements show clear underabundances in the range $-0.2$ to $-0.9$\,dex, except for Li, which is derived with an upper limit, only. The systematic underabundances are consistent with the remarkably deficient Fe abundance, [Fe/H]$=-0.71$ dex, most unusual among main-sequence A stars with very slow rotation, almost all of which are Am or Ap stars. Thus, the observed pattern is {\it not} characteristic of Am or Ap stars, where Fe-peak and rare-earth elements are overabundant (e.g., \citealt{Smith1971, Adelman1973}) -- often dramatically so.

The abundances of volatile elements, namely C and O, are also important for distinguishing metal-poor stars from the chemically peculiar class of $\lambda$\,Boo stars, which have Fe-peak elements that are underabundant, but volatile elements that are solar \citep{Stuerenburg1993}. Since we observe the volatile elements to be sub-solar in KIC\,11145123, we conclude it is not a $\lambda$\,Boo star, either.

The abundance pattern relative to Fe is shown in the bottom panel of 
Fig.\,\ref{Fig:abundances}, where enhancements of [X/Fe] $ \sim 0.1$ -- $0.5$ dex can be seen for all elements except Mn and Cr. 
The behaviour of [X/Fe] for $\alpha$ elements 
(C, O, Mg, Si, S, Ca, and Ti) resembles those well known among
metal-poor stars.
 Y, Zr, Ba and rare-earth elements show similar enhancements of $\sim 0.1$ --
$0.5$ dex. As established in the pattern of [X/H], the pattern of [X/Fe] 
again is inconsistent with characteristics of Am and Ap
stars. We discuss this further in sections \ref{sec:amstars} and \ref{sec:apstars} below.

\section{Model}
\label{sec:model}

It is a well-known problem that model frequencies in $\delta$~Sct and $\gamma$~Dor stars are difficult to match to the precision of the observed frequencies. Nevertheless, to proceed it is assumed that finding a ``best model'' -- one that has frequencies close to the observed ones -- is an informative procedure that gives some real information on the evolutionary state and structure of the star. In their discovery paper of the core-to-surface rotation of KIC\,11145123, \citet{Kurtz2014} tried to find asteroseismic stellar models that best fitted the pulsation frequencies -- particularly the g-mode series of frequencies and some of the p-mode frequencies. 

This suggested to \citet{Kurtz2014} that the star is near the terminal-age main-sequence, has enhanced helium abundance, and perhaps low metallicity. That led to the conjecture that KIC\,11145123 is a Population II star, which was the impetus for our acquiring the high-resolution Subaru spectrum and carrying out this study. As we have seen in Section\,\ref{sec:abundances}, KIC\,11145123 has very low metallicity, hence we have searched for models consistent with observed pulsation frequencies of KIC\,11145123 in the same way as described in \citet{Kurtz2014},  except that we have adopted here much lower metallicity, $Z = 0.003$, according to the spectroscopic analysis in this paper.

\citet{Kurtz2014} obtained rotationally split high-order dipole g-mode as well as p-mode frequencies from the {\it Kepler} photometric data. To constrain our models we used the mean period separation of the dipole g~modes, $\Delta P_{\rm g}$, (rather than each g-mode frequency) and  the frequencies of the five largest-amplitude p~modes; for rotational multiplets, the central frequencies (corresponding to zonal modes) were used. 

It is difficult to determine precisely the period spacing of the g~modes of KIC\,11145123 because this slightly depends on the range of periods and it has some irregularities (see figure 6 in \citealt{Kurtz2014}). We adopted $\Delta P_{\rm g} =0.0245 \pm 0.0002$\,d for the purpose of constraining our models, although the models obtained are not very sensitive to the exact value of $\Delta P_{\rm g}$. 

The largest-amplitude p~mode at a frequency of $17.964$\,d$^{-1}$ ($\nu_1$; we follow the frequency nomenclature of  \citealt{Kurtz2014}, where this is $\nu_1$, and so-on for further frequencies below) shows no rotational splitting -- it is a singlet; this is considered to be a radial mode. The other four large amplitude frequencies; 18.366\,d$^{-1}$ ($\nu_2$), 16.742\,d$^{-1}$ ($\nu_3$), 19.006\,d$^{-1}$ ($\nu_4$), and 22.002\,d$^{-1}$ ($\nu_5$) are triplets and quintuplets; i.e. nonradial modes of $\ell =1$ and $2$. We first obtained models having $\Delta P_{\rm g}$ and a radial mode consistent with $\nu_1$, then we examined the consistency with the other four p~modes by calculating the mean deviation of the frequencies, and by comparing $T_{\rm eff}$ and $\log g$ with the spectroscopic values of KIC\,11145123 derived in Section\,3 and also given in Table\,\ref{tab:01}.

Evolutionary models were calculated using Modules for Experiments in Stellar Evolution (MESA; \citealt{Paxton2013}) with the same settings as in Kurtz et al (2014): i.e., the heavy element abundance, $Z=0.003$, was scaled by the solar mixture of \citet{Asplund2009}; the radiative OPAL opacity tables \citep{iglesias&rogers1996} were used; and mixing length was set equal to 1.7\,$H_{\rm p}$, where $H_{\rm p}$ is the  pressure scale height. Atomic diffusion was activated in the code to erase noise in the Brunt--V\"ais\"al\"a frequency. The parameters specifying an evolutionary track are mass, initial chemical composition $(X,Z)$, and a core-overshooting parameter $h_{\rm ov}$, in which $h_{\rm ov}H_{\rm p}$ gives an exponential decay length of mixing efficiency \citep{herwig2000} above the convective core boundary. (Another often used overshooting parameter $\alpha_{\rm ov}$, in which a step-wise mixing in a range of $\alpha_{\rm ov}H_{\rm p}$ is assumed, approximately corresponds to $10h_{\rm ov}$.)

\begin{figure}
\centering
\includegraphics[width=0.9\linewidth,angle=0]{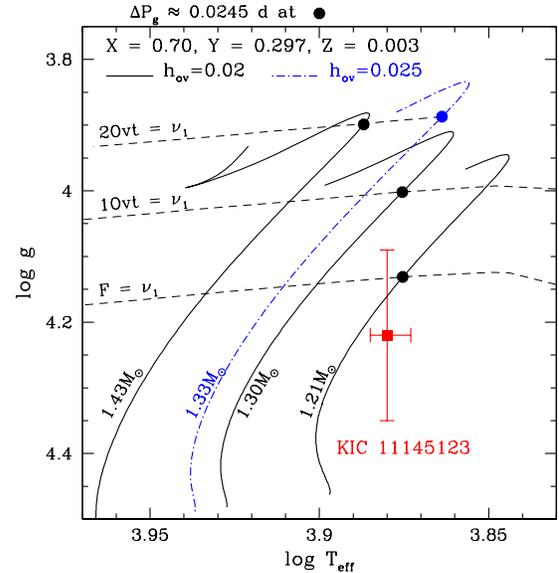}
\caption{Logarithmic effective temperature, $\log T_{\rm eff}$, versus surface gravity, $\log g$, diagram. Some main-sequence evolutionary tracks for a chemical composition of $(X,Z)= (0.7,0.003)$ are shown by solid and dash-dotted lines. The solid lines correspond to models with a core-overshooting parameter of $h_{\rm ov}=0.02 \approx \alpha_{\rm ov}/10$, while the dash-dotted line corresponds to $h_{\rm ov}=0.025$. Each horizontal dashed line indicates the locus where the largest-amplitude pulsation frequency of  KIC\,11145123 ($\nu_1 = 17.964$\,d$^{-1}$) is equal to the radial fundamental (F) mode, the first overtone (1Ovt) mode, or  the second overtone (2Ovt) mode frequency. A big dot on an evolutionary track indicates the position where the mean period spacing of high-order dipole g~modes is $\Delta P_{\rm g} = 0.0245$\,d (the observed value); $\Delta P_{\rm g}$ decreases with evolution (see, e.g., \citealt{Kurtz2014}). A horizontal dashed line does not necessary cross a track at a dot, although such tracks are selected in this diagram. The position of  KIC~11145123 with error bars is based on the spectroscopic results summarised in Table\,\ref{tab:01}. }
\label{fig:lgte_lgg}
\end{figure}

As evolution of the stellar model proceeds for a given set of parameters, the mean period spacing $\Delta P_{\rm g}$ decreases gradually, and at a particular evolutionary stage it becomes equal to the observed value 0.0245\,d,  shown as a big dot in Fig.\,\ref{fig:lgte_lgg} for four different evolutionary tracks. However, for most models no radial mode of the model reproduces $\nu_1$. When that was the case we changed the mass slightly and repeated the calculation until we obtained a model that reproduced both $\Delta P_{\rm g}$ and $\nu_1$. This corresponds, in Fig.\,\ref{fig:lgte_lgg}, to the positions of the filled circles, where the evolutionary tracks cross a horizontal dashed line. 

Since it is possible for $\nu_1$ to be identified as either the fundamental (F) mode, the first overtone (1Ovt) mode, or the second overtone (2Ovt) mode, we find, in general, three such models for a given set of  $(X,Z,h_{\rm ov})$. Fig.\,\ref{fig:lgte_lgg} shows the locations in the $\log T_{\rm eff}-\log g$ plane of three models with $(X,Z,h_{\rm ov})=(0.70,0.003,0.02)$ and one model with $(0.70, 0.003, 0.025)$. For a given $X$, larger $h_{\rm ov}$ yields a cooler model, so that we can tune $h_{\rm ov}$ to have a model with $T_{\rm eff}$ consistent with the spectroscopic value of KIC\,11145123.

\begin{figure}
\centering
\includegraphics[width=0.95\linewidth,angle=0]{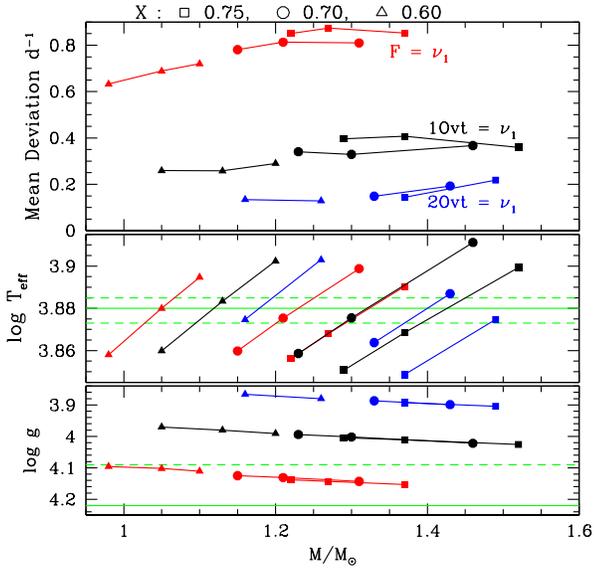}
\caption{The surface gravity (bottom panel), the effective temperature (middle panel), and the mean deviation from observed p-mode frequencies (top panel) of each model satisfying the conditions of $\Delta P_{\rm g} \approx 0.0245$\,d and a radial mode frequency equal to $\nu_1$ (i.e., dots crossing one of the horizontal dashed lines in Fig.~\ref{fig:lgte_lgg}) are plotted.  Different symbols are used for models with different initial hydrogen abundances ($X = 0.75, 0.70, 0.60$). For a given $X$, some values for the overshooting parameter $h_{\rm ov}$ in a range between $0.01$ and $0.03$ are adopted to obtain models with $T_{\rm eff}$ close to the spectroscopic value of KIC\,11145123. Models with the same $X$ but different $h_{\rm ov}$ are connected by lines. Red, black, and blue colours are used for models which identify the observed $\nu_1$ as the fundamental  (F) mode, the first overtone (1Ovt) mode, or the second overtone (2Ovt) mode, respectively (see top panel). The spectroscopic values of  $\log T_{\rm eff}$ and $\log g$ and their error ranges of KIC\,11145123 are shown by horizontal lines.}
\label{fig:meandev}
\end{figure}

Mean deviations of $\nu_2 \dots \nu_5$ of  KIC~11145123 from the models thus obtained  are shown in Fig.~\ref{fig:meandev}, in which $T_{\rm eff}$ and $\log g$ are also compared with those of  KIC\,11145123. The mean deviations for the models with $\nu_1$ being identified as second radial overtone modes (2Ovt)  are in most cases less than 0.2\,d$^{-1}$ (i.e. less than $\sim$1\,per~cent); these are significantly smaller than those of the models with $\nu_1$ being identified as the fundamental radial overtone (F). However, the surface gravities of the models that identify $\nu_1$ as belonging to the second radial overtone (2Ovt) are lower than the spectroscopically measured $\log g=4.22\pm0.13$ (Table\,\ref{tab:01}) by about 2.4$\sigma$. On the other hand, although models that identify $\nu_1$ with the fundamental radial mode (F) have surface gravities more consistent with the spectroscopic value, the mean deviations of the pulsation frequencies are much higher ($\sim$5\,per~cent). It is not clear at present how to tune the models for better consistency with the observations. It seems probable that the internal structure of KIC\,11145123 deviates considerably from the structure of the models calculated assuming single-star evolution; further fine-tuning of the models needs a dedicated study of the internal structure of non-standard models.  

We note that  models with a larger initial hydrogen abundance are slightly more massive, but the mean deviations hardly depend on $X$.  A hydrogen abundance of $X=0.75$ (corresponding to $Y=0.247$) is probably close to normal for a $Z=0.003$ single star. However, the ages of the $X=0.75$ models are  $\sim 1.5 - 3\times 10^9$~yr; which are too young for a single star, having evolved normally in our Galaxy, to have the low metallicity of $Z = 0.003$ ([Fe/H] = $-$0.7) of KIC\,11145123.

\section{Discussion}
\label{sec:discussion}

To clarify the properties of our target, we discuss the atmospheric parameters, the rotation velocity and the abundance pattern in this section. We discuss interpretations of these and the asteroseismic results in terms of a blue straggler.

\subsection{Atmospheric parameters}
\label{sec:atparam}

We determined the atmospheric parameters of $T_{\rm eff}$, $\log g$, $\xi$, and [Fe/H] spectroscopically, using Fe~{\sc i} and Fe~{\sc ii} lines. Values of $T_{\rm eff}$, $\log g$ and [Fe/H] are also available from broadband photometry and colour indices from the automated analysis of \citet{Huber2014}. There are differences between our values and those of \citet{Huber2014}, which are $\Delta$\Teff$=-461$ K, $\Delta$\logg$=+0.25$, and $\Delta$[Fe/H]$=-0.57$ dex, in the sense [(ours) $-$ (Huber et al.)].

While we fully expect that our atmospheric parameters from our high resolution, high S/N spectra are more accurate than those from much lower resolution broadband photometry, to examine the validity of our derived parameters further, we compare them with those of 117  {\it Kepler} A and F stars investigated by \citet{Niemczura2015}, who obtained the final values of these parameters using  Fe\,{\sc i} and Fe\,{\sc ii} lines  based on high-resolution spectra, in the same manner as we did. They compared the behaviour of the parameters between their values and those of \citet{Huber2014}, taking differences of each parameter into account.

For $T_{\rm eff}$, figure 3(a) of \citet{Niemczura2015} depicts the behaviour of $\Delta T_{\rm eff}  = T_{\rm eff} {\rm (SPEC)} - T_{\rm eff} {\rm (H2014)}$ [where $T_{\rm eff}$(SPEC) is their value and $T_{\rm eff}$(H2014) that of \citet{Huber2014}], which has large scatter of $\pm 600$\,K as a function of \Teff(SPEC). When we plot our $\Delta$\Teff$=-461$\,K in their figure 3(a), it is included in the negative scatter range. \citet{Niemczura2015} suggested that the accuracy of their spectroscopic temperatures is supported because the temperatures derived from spectral energy distributions agree well with \Teff(SPEC) within a scatter of about $\pm 300$\,K. According to their suggestion, and from an evaluation of the quality of each method, we consider our atmospheric parameters, \Teff, $\log g$ and [Fe/H] to be more reliable than those from broadband photometry. This is unsurprising. 

The values of microturbulence $\xi$ derived by \citet{Niemczura2015} are plotted in their figure 5. Our $\xi$ is consistent with the behaviour of non-chemically peculiar stars at \Teff$\sim 7600$\,K.

\subsection{Rotation velocity and magnetic braking; Ap stars}
\label{sec:rotvel}

Our apparent projected rotation velocity, $v_{\rm a} \sin i  = 5.9 \pm 0.2$\,km~s$^{-1}$ is extremely slow compared with the distribution of $v \sin{i}$ for the stars with similar temperatures in the range  $7000 \le T_{\rm eff}  \le 8000$\,K \citep{Niemczura2015}. Their figure 7(e) shows, and their electronic table 3 lists, values of $v \sin{i}$ distributed between 8 and 260\,km\,s$^{-1}$, but mostly between 50 and 150\,km\,s$^{-1}$ (note that they do not take into account macroturbulence, hence their minimum $v \sin i$ is overestimated). Thus the reason why KIC\,11145123 rotates so slowly is unclear; this important point needs to be explored: Why does KIC\,11145123 rotate so slowly compared with the majority of A and early F stars?

It is well known that Ap stars with magnetic fields rotate more slowly than normal A stars. \citet{Stepien2000} suggested
that magnetic Ap stars become slow rotators due to loss of angular momentum by magnetic braking in the pre-main-sequence phase. While our search for a magnetic field in KIC\,11145123 yielded a null result, it did not rule out a field strength of the order of 1\,kG, which is typical of the field strengths found in many Ap stars. However, our abundance analysis clearly shows that KIC\,11145123 is not an Ap star -- as we argue in section \ref{sec:apstars} -- hence we take the null result for the magnetic measurement to be correct, and conclude that magnetic braking is not a viable hypothesis for the remarkably slow rotation of this star.

\subsection{Phase modulation and a stringent upper limit on the mass of any companion; Am stars}
\label{sec:phasemod}

\begin{figure}
\begin{center}
\includegraphics[width=0.48\textwidth]{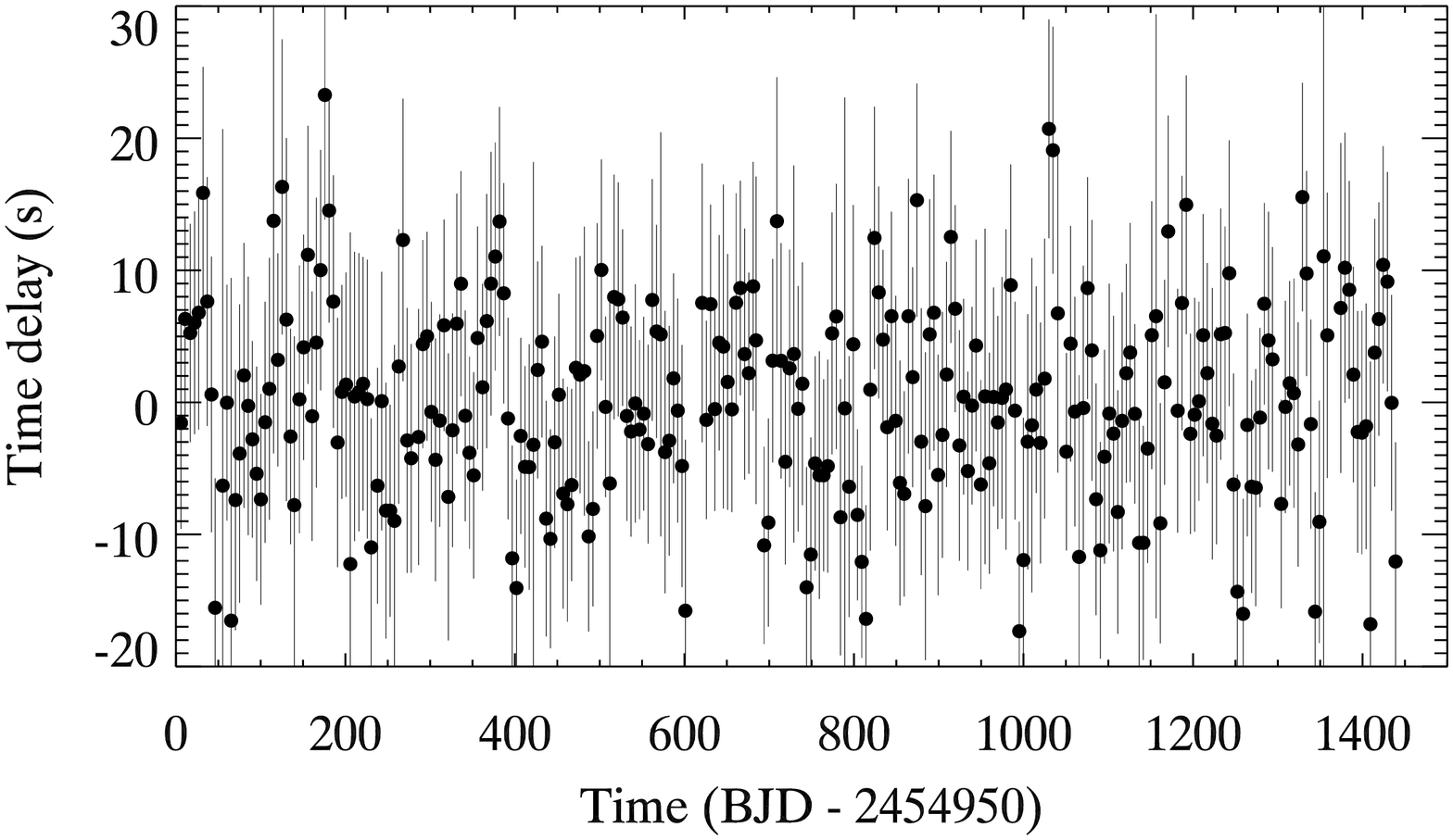}
\includegraphics[width=0.48\textwidth]{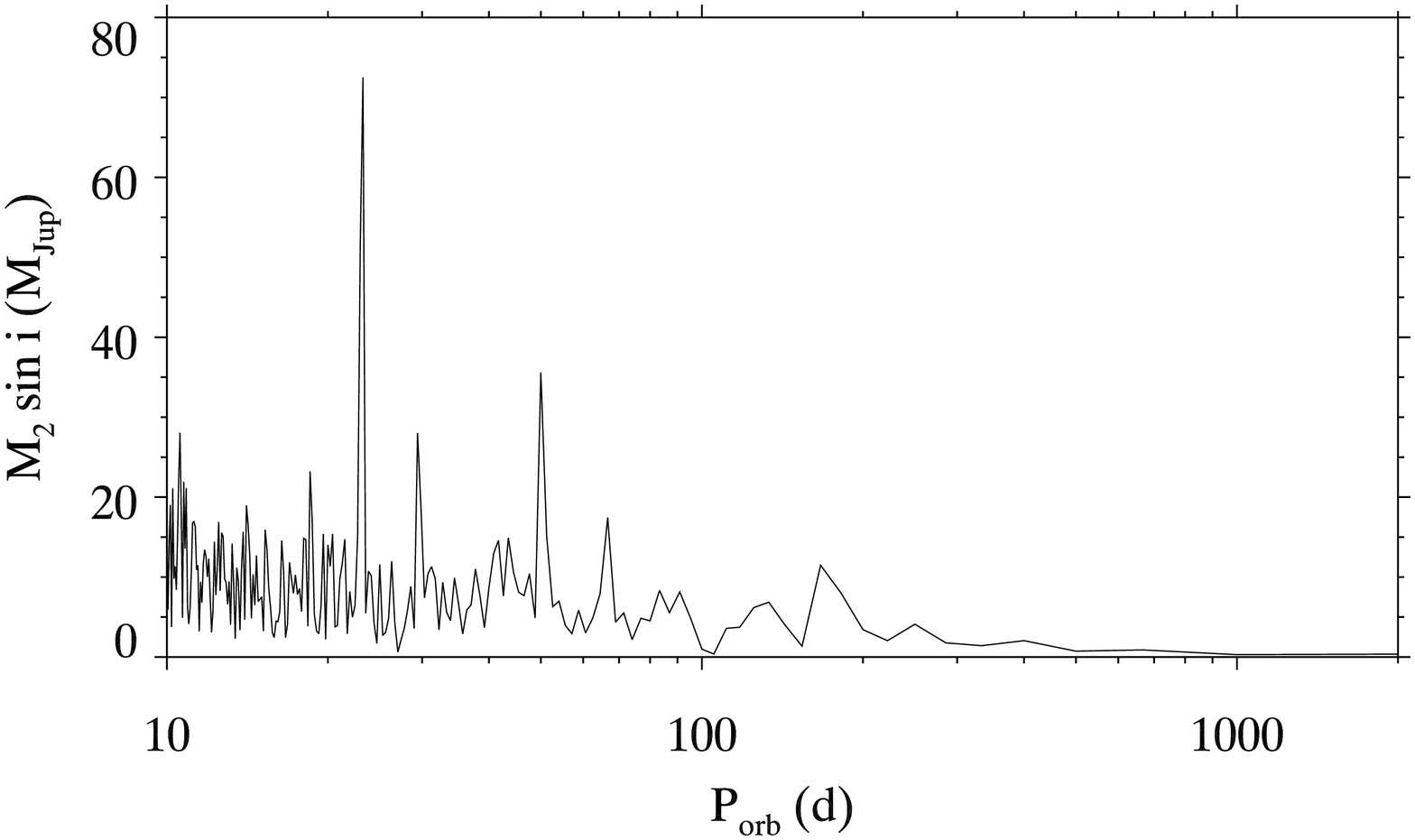}
\caption{Top: Time delays derived from the p-mode singlet. Bottom: upper limits on $M_2\sin i$ from Fourier analysis of the time delays. }
\label{fig:pm}
\end{center}
\end{figure}

It is well known that metallic-lined A stars -- the Am stars -- are relatively slow rotators with $v \sin i \le 100$\,km\,s$^{-1}$ \citep{abt1967}. These stars are typically in close binary systems with orbital periods of $1 - 10$\,d, where spin-orbit synchronisation has slowed the rotation of the Am star. This slow rotation is a requisite for the effects of atomic diffusion to alter the atmospheric abundances in Am stars. We therefore searched for evidence of a binary companion to KIC\,11145123.  

We found no evidence of such a companion, and have derived stringent upper limits for the allowable mass of any possible companion using the phase modulation (PM) method \citep{murphyetal2014,murphy-shibahashi2015}. We divided the data into 5-d segments to search for changes in the pulsation phase that could be caused by binary motion. Phase changes were converted into time delays, which are shown in the top panel of Fig.\,\ref{fig:pm}. We used only the p-mode singlet; the g\:modes and non-radial p\:modes are rotationally split, and the azimuthal components are not resolved in 5-d light-curve segments. To improve the precision we pre-whitened all other peaks in the Fourier transform of the light curve with amplitudes exceeding 0.2\,mmag. The lower panel of Fig.\,\ref{fig:pm} shows the upper limits to $M_2\sin i$ in the 10--2000\,d period range. A companion with $M_2\sin i$ consistent with a brown dwarf could exist at some periods, but the peaks in the lower panel of Fig.\,\ref{fig:pm} could be caused by residual low-amplitude oscillations near the p-mode singlet instead. 

We thus find that no stellar mass companion is present for orbital periods of $10-2000$\,d from our PM analysis, and shorter periods of $1 - 10$\,d would produce ellipsoidal variation in the high-precision Kepler light curve. We certainly rule out rotation and orbital coupling in a 100-d binary. The caveat is that this is dependent on the orbital inclination; if there were a companion in an orbit with very low inclination, then we would be insensitive to that, both in the PM analysis and in the ellipsoidal variations. However, \citet{Kurtz2014} showed that the g-mode pulsations have very low visibility of the axisymmetric modes, which strongly suggests that the modes have high inclination. Unless there is a strong misalignment of the purported orbital inclination and the pulsation axis, this also argues against any companion. Finally, no F, G or K main-sequence companion can be present because we would see evidence in our high resolution spectra. 

With our abundances -- discussed in detail in the next section -- and these arguments, we conclude that KIC\,11145123 is not an Am star and that it is {\it currently} not in a binary system. Thus orbital synchronisation is not the cause of the slow rotation. However, we return to this latter conclusion in the discussion below -- and this is the reason we italicised ``{\it currently}'' -- because we conjecture that previously KIC\,11145123 may have been in a binary system to give it its current blue straggler stellar structure. 

\subsection{Abundance pattern}
\label{sec:abunpattern}

In this Section we compare the abundance patterns relative to Fe in our target and in Am, Ap and blue stragglers to 
clarify the properties of our target and to give an answer to the prediction by \citet{Kurtz2014} that our target is either an Am star or a blue straggler.

\subsubsection{Comparison with Am stars}
\label{sec:amstars}

\begin{figure}
\centering
\includegraphics[width=0.9\columnwidth]{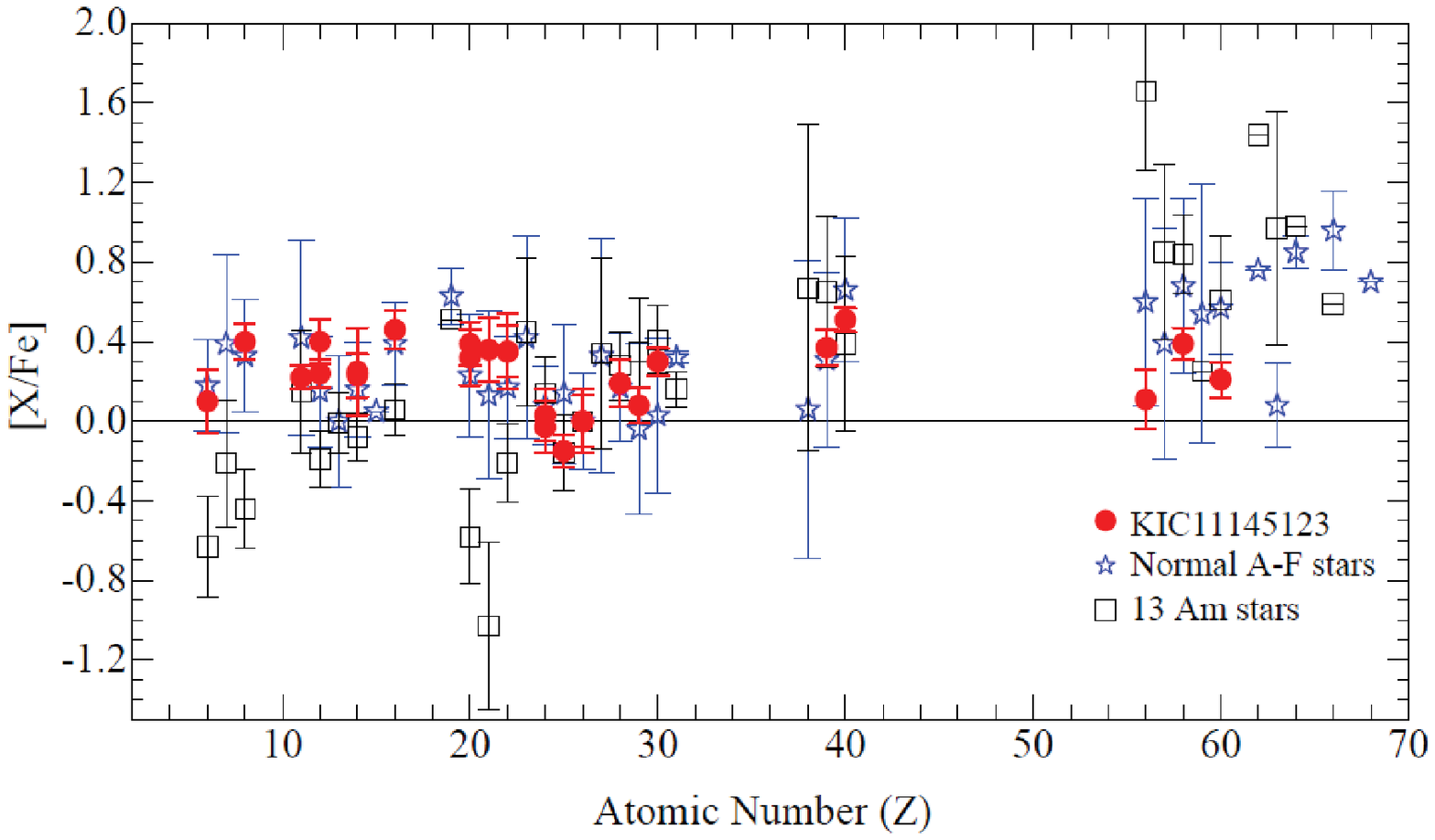}
\includegraphics[width=0.9\columnwidth]{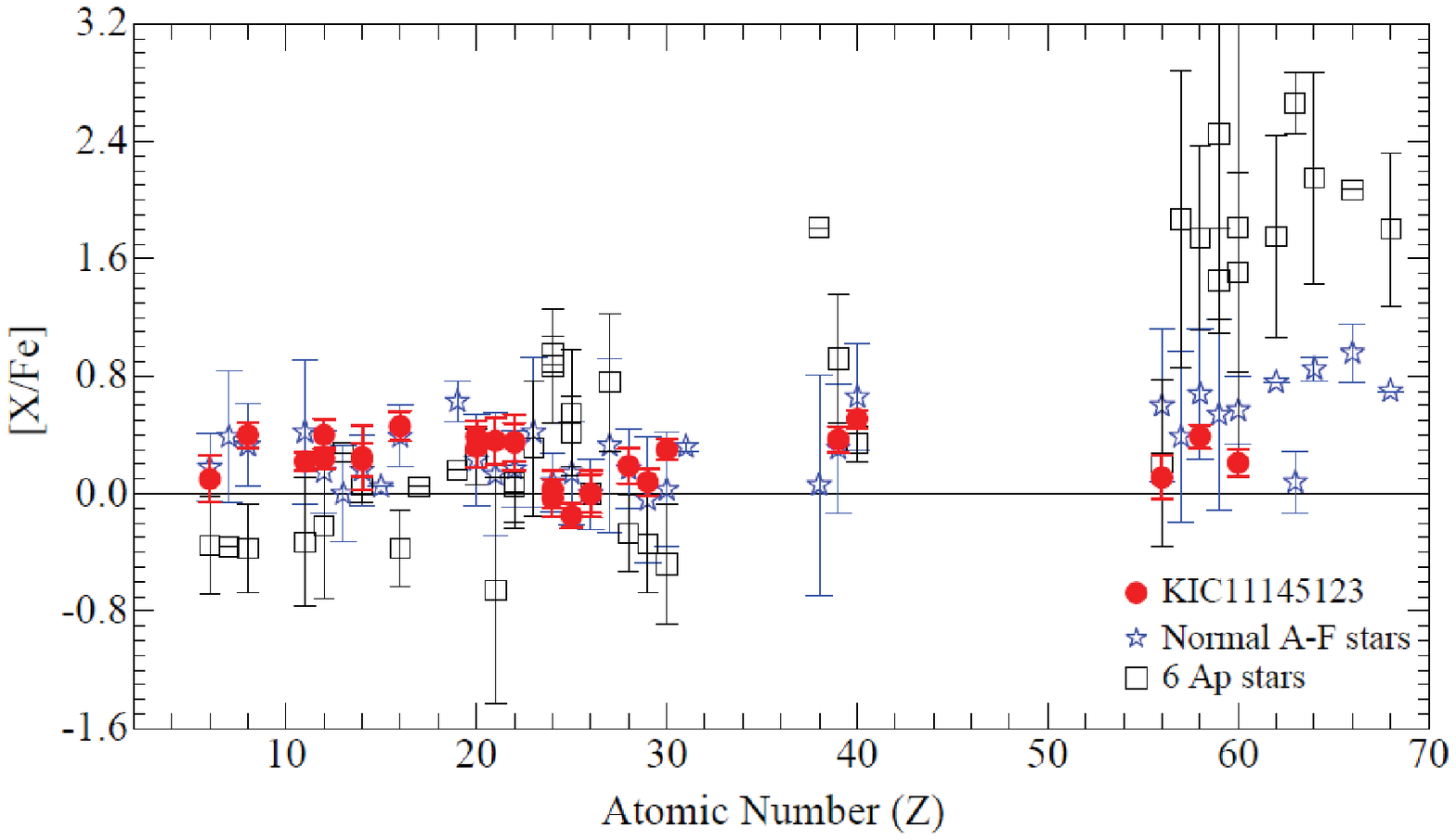}
\caption{Top panel: Comparison of abundance patterns between our target and normal A/F and Am stars. Bottom panel: Comparison of abundance patterns between our target and normal A/F stars, and the average abundance pattern of five roAp stars and one $\delta$ Sct star with an Ap chemical signature. From these comparisons we conclude that KIC\,11145123 is neither an Am star nor an Ap star.}
\label{Fig:comparison}
\end{figure}

In the top panel of Fig.\,\ref{Fig:comparison} we compare the abundance pattern (except for Li) of KIC\,11145123 with those of  the average abundances of the 96  normal A--F stars and 13 Am stars adopted from \citet{Niemczura2015}.

The abundance pattern of our target is consistent with that of normal A--F stars, and inconsistent with that of Am stars because the typical underabundances of Ca and Sc, and overabundances of Ba and rare-earth elements, for Am stars are not observed. Hence we conclude that our target is not an Am star.

\subsubsection{Comparison with roAp stars}
\label{sec:apstars}

In the bottom panel of Fig.\,\ref{Fig:comparison} we compare the abundance pattern of our target with those of the same normal A--F stars as adopted above, and the average abundance pattern of five roAp stars: HD 203932,  \citet{Gelbmann1997}; 10 Aql,  $\beta$ CrB, and 33 Lib,  \citet{Ryabchikova2004};  KIC 4768731, \citet{Niemczura2015}; and one $\delta$ Sct star with an Ap chemical signature:  HD 41641,  \citet{Escorza2016}.

The abundance pattern of our target is not consistent with those of roAp stars: C and O are not underabundant, and rare-earth elements show no enhancement. The abundance pattern is consistent with that of normal A--F stars, just as 
in the comparison with Am stars. Hence we conclude that our target is not an Ap star, rapidly oscillating or otherwise.

\subsubsection{Comparison with $\lambda$\,Boo stars}
\label{sec:lboostars}

\begin{figure}
\centering
\includegraphics[width=0.9\columnwidth]{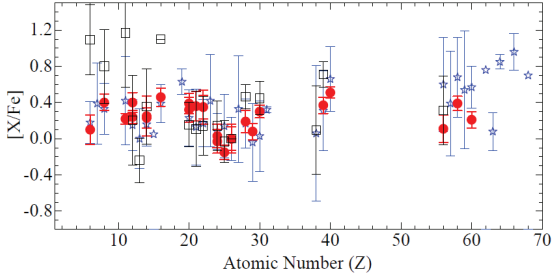}
\caption{Comparison of abundance patterns between our target (filled circles) and normal A/F stars (stars) and 25 $\lambda$\,Bootis stars (open squares). The systematic enhancement of volatile elements of C, O, Na and S that is observed in $\lambda$\,Boo stars is  not seen in our target star. KIC\,11145123 is not a $\lambda$\,Boo star.}
\label{Fig:lboo}
\end{figure}

Fig.\,\ref{Fig:lboo} compares the abundance patterns of KIC\,11145123, normal A/F stars  and 25 $\lambda$\,Bootis stars. The sample of $\lambda$\,Boo stars consists of nine, six, and ten stars adopted from \citet{Stuerenburg1993}, \citet{paunzen1999} and \citet{heiter2002}, respectively, which are confirmed as $\lambda$\,Boo stars by \citet{murphy2015}, and their average abundances are plotted in Fig.\,\ref{Fig:lboo}. The volatile elements  C, O, Na and S show enhancement of $\sim$1\,dex compared to Fe, in contrast to KIC\,11145123 and also to normal A/F stars.  We therefore conclude that KIC\,11145123 is not a $\lambda$\,Boo star.

\subsubsection{Comparison with blue stragglers}
\label{sec:bluestragglers}

KIC\,11145123 has a large radial velocity of $V_{\rm  h} = - 135.4
\pm 0.2$\,km\,s$^{-1}$. We do not have a parallax, therefore no distance
measurement, but can get a rough estimate of distance from the
luminosity of the best asteroseismic models of \citet{Kurtz2014}, who
give $\log L/{\rm L_{\odot}} = 1.1$, and the approximation that the {\it
Kepler} magnitude, Kp = 13.1, is close to Johnson $V$. Together those
give a distance of about 1.7\,kpc. The galactic longitude of
KIC\,11145123 is $81.7^{\circ}$, which means the star is nearly in the
direction of galactic rotation. With the approximate distance, it is
easy to calculate that KIC\,11145123 is at about the same distance from
the galactic centre as is the Sun. 
If we assume that the observed radial velocity of KIC\,11145123 is close
to the component of its peculiar velocity along the direction of
Galactic rotation, given the Sun's galactic orbital velocity of
220\,km~s$^{-1}$ and its component of the peculiar velocity $V_{\odot}
=$ 5.2\,km~s$^{-1}$, the large velocity of approach  of KIC\,11145123 of
$-135$\,km~s$^{-1}$ shows it, therefore, to be orbiting in the direction
of rotation of the Galaxy with the rotation component of  $\Theta =$
90\,km~s$^{-1}$, -- albeit much more slowly than the Sun and typical
Population I stars -- indicating that it is a halo, or runaway star.
 The galactic latitude of KIC\,11145123 is $12.5^{\circ}$, which at a nominal distance of 2\,kpc puts it about 400\,pc above the galactic plane, near the edge of the thin disk, of which it is probably not a member. Without a parallax and proper motion, we cannot characterise the space motions more than this at present, but these properties, along with the low metallicity, are shared with the blue stragglers that are known to exist in many globular clusters, old open clusters, and among the field Population II stars in the Galactic bulge and halo \citep[e.g.][]{Glaspey1994, Lovisi2013}.

\begin{figure}
\centering
\includegraphics[width=0.9\columnwidth]{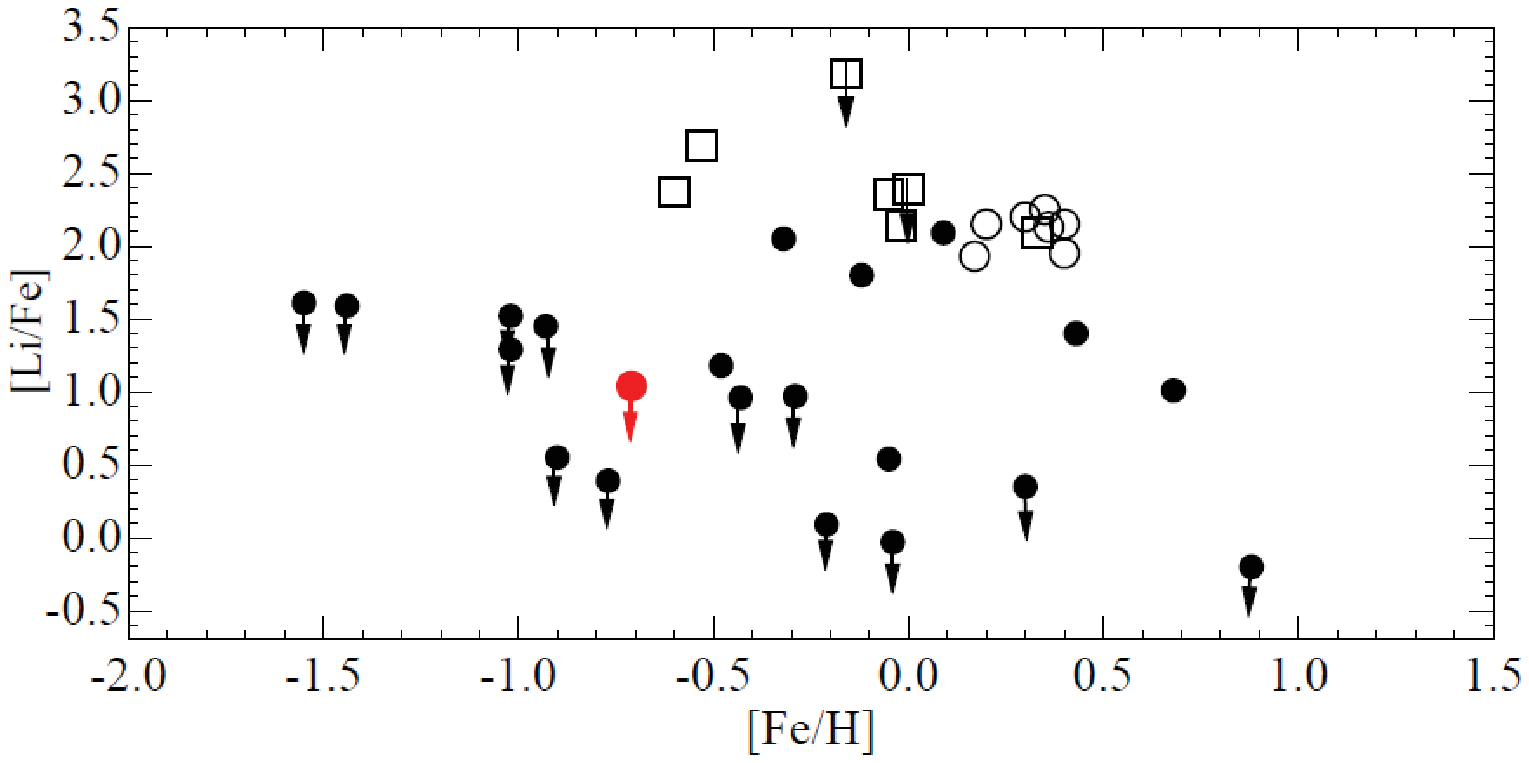}
\caption{Behaviour of [Li/Fe] vs. [Fe/H] for our target (red filled circle), 20 high  velocity A--F stars \citep[][black filled
 circles]{Glaspey1994}, and normal A stars (\citealt{Takeda2012}, open squares; \citealt{Burkhart2000}, open circles).}
\label{Fig:bs_abundances}
\end{figure}

As a final discussion on the abundance pattern of our target, we compare our target with  blue stragglers in the abundances of Li, C, and O. In Fig.\,\ref{Fig:bs_abundances} we depict the behaviour of [Li/Fe] as a function of [Fe/H] for the 20 high velocity A--F stars taken from \citet[][Table 3]{Glaspey1994} and our target. They identified the 15 Li-weak stars as halo and thick-disk blue stragglers, or old and mostly thin-disk blue stragglers, of which 13 stars  are shown with upper limits on the Li abundance.  We also plot [Li/Fe] of 14 normal A stars taken from \citet{Takeda2012} and \citet{Burkhart2000}.

As can be seen from Fig.\,\ref{Fig:bs_abundances}, the [Li/Fe] of our target is certainly lower than those of normal A stars and a few Li strong stars of \citet{Glaspey1994}, but is compatible with the Li-weak blue stragglers. The low abundance of Li on its own argues that KIC\,11145123 during its lifetime has had a thick surface convection zone that took $^7$Li down to levels where the temperature was $2.4$\,MK, where it is destroyed in fusion reactions. Main-sequence A and early F stars do not have convection zones thick enough to do this. We therefore deduce that KIC\,11145123 had an earlier main-sequence, or post-main-sequence, life with a convection zone thick enough to reach the temperature of Li burning.

\begin{figure}
\centering
\includegraphics[width=0.9\columnwidth]{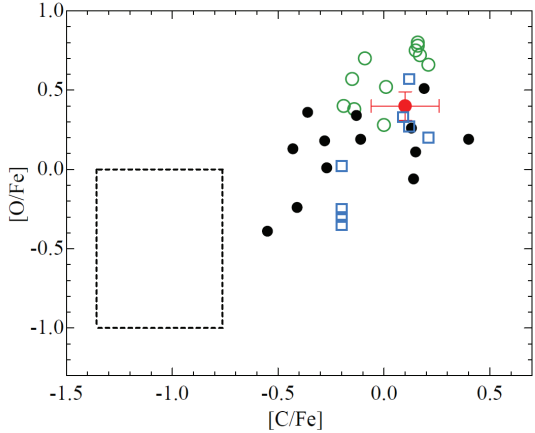}
\caption{Behaviour of [O/Fe] vs [C/Fe]  for the blue stragglers in the  globular clusters NGC\:6752 (\citealt{Lovisi2013}, black filled circle) and M\:4 (\citealt{Lovisi2010}, green open circle), and for turn-off dwarf stars in NGC\:6752 \citealt{Carretta2005}, blue open square). Our target is shown with red filled circle. The distribution of CO-depleted blue stragglers \citep{Ferraro2006} is shown as the box with a dashed line.} 
\label{Fig:CO_plot}
\end{figure}

In Fig.\,\ref{Fig:CO_plot} we illustrate the behaviour of  [O/Fe] vs [C/Fe]  for our target and the blue stragglers
in the globular clusters of  NGC\:6752 and M\:4 adopted from \citet{Lovisi2013} and \citet{Lovisi2010}, respectively.
The turn-off dwarf stars in NGC\:6752 adopted from \citet{Carretta2005}  and the distribution region of the 
CO-depleted blue stragglers in 47 Tuc adopted from \citet{Ferraro2006} are shown for comparison.
 
The adopted sample of blue stragglers in NGC\:6752 has $\log g \sim 4$ and $v \sin{i} < 40$\,km\,s$^{-1}$ -- consistent with those of our target -- while [Fe/H] ($\sim -1.5$\,dex) is different. The sample of blue stragglers in M4 shows $v \sin{i} < 5$ \kms\ in most cases, but is different in [Fe/H] ($\sim -1.3$\,dex) and radial velocities ($+70$ to $+$90 \kms). The turn-off dwarf stars in NGC\:6752 overlap with these blue stragglers. 
These blue stragglers and turn-off dwarf stars are clearly distributed in a different region outside the region of  CO-depleted blue stragglers in 47 Tuc. Our target's location is compatible with the observed abundances of both the blue stragglers and the turn-off dwarf stars. \citet{Lovisi2010} and \citet{Lovisi2013}  found that there is no depletion of C and O abundances in their samples of blue stragglers with slow rotation, and suggested that the studied blue stragglers in M4 and NGC\:6752 may derive from stellar collisions, for which no chemical anomalies are expected.

Taking into account the consistencies found in the behaviour of Li, C, and O between our target and the blue stragglers, 
we conclude that our target is a blue straggler. 

\subsection{Summary of the spectroscopic analysis}
\label{sec:summary}

We analysed the properties of our target, KIC\,11145123, based on high-dispersion spectra obtained by HDS on the Subaru telescope. We summarise the results of the spectroscopic analysis of our target as follows:

(1) The atmospheric parameters are derived  from the Fe~{\sc i} and Fe~{\sc ii} lines:  $T_{\rm eff} = 7590 ^{+80}_{-140}$\,K, 
$\log g = 4.22 \pm 0.13$, $\xi =3.1\pm 0.5$\,km~s$^{-1}$, and $[{\rm Fe/H}]= -0.71 \pm 0.11$. The $T_{\rm eff}$ and [Fe/H] values are more reliable than the values derived from photometric colours given in the KIC. The low [Fe/H] value is unusual among late-A stars, implying that our target is a Population II star.

(2) The apparent projected rotation velocity $v_{\rm a} \sin{i} = 5.9 \pm 0.2$\,km\,s$^{-1}$ is consistent with the asteroseismically predicted $v_{\rm eq} = 1$\,km~s$^{-1}$ convolved with macoturbulent and pulsational broadening. 
 
(3) The comparisons of abundance patterns of [X/Fe] between our target, normal A--F stars, Am stars,  and roAp stars show that our target is neither an Am star nor  an Ap star, but follows consistently the pattern of normal A--F stars.

(4) Our analysis of [Li/Fe], [C/Fe], and [O/Fe] abundances is consistent with our target being a blue straggler.
This suggestion is consistent with the low [Fe/H], the slow rotation, and the high radial velocity of our target, which are
well known to be observed in Population II stars.

\section{Conclusion: The blue straggler hypothesis }
\label{sec:blue straggler}

KIC\,11145123 presents a problem. It is an exceedingly slowly rotating A star with a rotation period of 100\,d, and is neither an Am or Ap star, so cannot have lost angular momentum as those stars do. How did this star come to be so slowly rotating? This problem is not an isolated one for KIC\,11145123. Other {\it Kepler} A stars that have had interior and surface rotation measured by asteroseismology also show unusually slow rotation \citep{Saio2015,Schmid2016,Murphy2016}. It is true that there is a selection effect in these first studies of core-to-surface rotation in that the rotational splitting in the nonradial mode multiplets is easier to see for slow rotation. Nevertheless, the slow rotation is real, and needs explanation. 

We have found in this paper that KIC\,11145123 has the attributes of a field blue straggler. It has low metallicity of $\rm{[Fe/H]} = -0.71$. This alone probably requires that star to be older than the Sun, since it had to be born at a time and place in galactic history when the metallicity of the interstellar medium was less than that when and where the Sun was born. Since this is variable with position within the Milky Way, no precise age can be ascribed to KIC\,11145123, but it certainly is much older than the $\sim$1\,Gyr lifetime of a main-sequence A star, hence it cannot have been born an A star. This is supported by our asteroseismic mass of 1.4\,M$_\odot$ for our best model of the star; this is much lower than the 1.7\,M$_\odot$ expected for a Population I A star of the same $T_{\rm eff} = 7600$\,K and $\log g = 4.2$. And the radial velocity of $-135$\,km~s$^{-1}$ shows the star to be Population II. 

There is a time and age problem. If KIC\,11145123 was born $5 - 10$\,Gyr ago, which would be consistent with its [Fe/H] = $-0.71$, then what has it been doing all this time to now be an A star with only a 1-Gyr main-sequence lifetime? We suggest that only merger or mass transfer in a binary system can account for this, and those are a common pathways for stars to become blue stragglers. Yet both pathways present problems. 

If we imagine a stellar merger, then KIC\,11145123 would have started life as a much lower mass star. It would have been below the Kraft Break, hence would have had relatively slow rotation to begin with. The merger would have increased its mass to the current 1.4\,M$_\odot$ of our best asteroseismic model, mixed the interior, and moved the star to the main-sequence with a somewhat enhanced He abundance, as in our best model. Mergers have been modelled with the formation of a circumstellar disk coupled to a $\sim$200\,G magnetic field to explain slow rotation of blue stragglers \citep{sills2005}, but those models do not specifically address the excessively slow rotation with a period of 100\,d. As we have pointed out, KIC\,11145123 is far from the only A--F star in the {\it Kepler} data that shows excessively slow rotation. A common mechanism is needed to understand the rotation, hence for the merger model to apply in this case, it must be a common pathway to the formation of blue stragglers, rather than a rare one. 

The other pathway to a blue straggler is for mass to be stripped by a companion in the evolution of a binary system. This, too, could leave the donor star with enhanced He abundance, and the change in structure with the mass loss could result in some internal mixing to get the resultant star back onto the main-sequence with its new mass and structure. Mass loss could also shed angular momentum. This seems a more promising idea than merger. But, we have put stringent upper limits that the mass of any current companion must be less than stellar, if we accept that the amplitudes of the g-mode dipole triplets give some information on the rotation axis, and that there isn't an extreme -- say nearly 90$^\circ$ -- spin-orbit misalignment in the purported binary. That leaves some room to manoeuvre in this scenario, but it, too, seems improbable, and we are looking for a mechanism that can produce many late A and early F very slowly rotating stars. 

We do not have a suggestion that is consistent with all that we have learned of this star from our spectroscopic and asteroseismic studies. The star is Population II.  It is metal poor and has high space velocity. It is not an Am or Ap star. It is on the main-sequence. It does have a 100-d rotation period. What is it and how did it get that way? We do not yet know. GAIA will provide a parallax and proper motion that will give better understanding of the star's space motion, but that is unlikely to give a definitive understanding of the evolutionary pathway to such an unusual object. Progress is more likely to come with the analysis of similar very slowly rotating A stars from existing {\it Kepler} data, and from data from the upcoming TESS mission.

\section*{Acknowledgements}

This work was partially supported by JSPS Grant-in-Aid for Scientific
Research (C) (Grant Number 16K05288). DWK thanks the Heiwa Nakajima
Foundation for generous support for a research visit to the University
of Tokyo. We thank the referee, F. Royer, for his constructive comments
which improved our paper.

\bibliographystyle{mn2e}
\bibliography{kic11145123_spectroscopy_11}

\begin{thebibliography}{48}
\expandafter\ifx\csname natexlab\endcsname\relax\def\natexlab#1{#1}\fi

\bibitem[{{Abt}(1967)}]{abt1967}
{Abt} H.~A., 1967, in Magnetic and Related Stars, {R.~C.~Cameron}, ed., pp.
  173--+

\bibitem[{{Adelman}(1973)}]{Adelman1973}
{Adelman} S.~J., 1973, \apj, 183, 95

\bibitem[{{Asplund} {et~al}\mbox{.}(2009){Asplund}, {Grevesse}, {Sauval}, \&
  {Scott}}]{Asplund2009}
{Asplund} M., {Grevesse} N., {Sauval} A.~J., {Scott} P., 2009, \araa, 47, 481

\bibitem[{{Burkhart} \& {Coupry}(2000)}]{Burkhart2000}
{Burkhart} C., {Coupry} M.~F., 2000, \aap, 354, 216

\bibitem[{{Carretta} {et~al}\mbox{.}(2005){Carretta}, {Gratton}, {Lucatello},
  {Bragaglia}, \& {Bonifacio}}]{Carretta2005}
{Carretta} E., {Gratton} R.~G., {Lucatello} S., {Bragaglia} A., {Bonifacio} P.,
  2005, \aap, 433, 597

\bibitem[{{Castelli} \& {Hubrig}(2004)}]{Castelli2004}
{Castelli} F., {Hubrig} S., 2004, \aap, 425, 263

\bibitem[{{Cayrel}(1988)}]{Cayrel1988}
{Cayrel} R., 1988, in IAU Symposium, Vol. 132, The Impact of Very High S/N
  Spectroscopy on Stellar Physics, {Cayrel de Strobel} G., {Spite} M., eds., p.
  345

\bibitem[{{Escorza} {et~al}\mbox{.}(2016){Escorza}, {Zwintz}, {Tkachenko}, {Van
  Reeth}, {Ryabchikova}, {Neiner}, {Poretti}, {Rainer}, {Michel}, {Baglin}, \&
  {Aerts}}]{Escorza2016}
{Escorza} A. {et~al.}, 2016, \aap, 588, A71

\bibitem[{{Ferraro} {et~al}\mbox{.}(2006){Ferraro}, {Sabbi}, {Gratton},
  {Piotto}, {Lanzoni}, {Carretta}, {Rood}, {Sills}, {Fusi Pecci}, {Moehler},
  {Beccari}, {Lucatello}, \& {Compagni}}]{Ferraro2006}
{Ferraro} F.~R. {et~al.}, 2006, \apjl, 647, L53

\bibitem[{{Gelbmann} {et~al}\mbox{.}(1997){Gelbmann}, {Kupka}, {Weiss}, \&
  {Mathys}}]{Gelbmann1997}
{Gelbmann} M., {Kupka} F., {Weiss} W.~W., {Mathys} G., 1997, \aap, 319, 630

\bibitem[{{Gizon} {et~al}\mbox{.}(2016){Gizon}, {Sekii}, {Takata}, {Kurtz},
  {Shibahashi}, {Bazot}, {Benomar}, {Birch}, \& {Sreenivasan}}]{gizon2016}
{Gizon} L. {et~al.}, 2016, Science Advances, 2, e1601777

\bibitem[{{Glaspey}, {Pritchet} \& {Stetson}(1994){Glaspey}, {Pritchet}, \&
  {Stetson}}]{Glaspey1994}
{Glaspey} J.~W., {Pritchet} C.~J., {Stetson} P.~B., 1994, \aj, 108, 271

\bibitem[{{Gray}(2014)}]{Gray2014}
{Gray} D.~F., 2014, \aj, 147, 81

\bibitem[{{Heiter}(2002)}]{heiter2002}
{Heiter} U., 2002, \aap, 381, 959

\bibitem[{{Herwig}(2000)}]{herwig2000}
{Herwig} F., 2000, \aap, 360, 952

\bibitem[{{Huber} {et~al}\mbox{.}(2014){Huber}, {Silva Aguirre}, {Matthews},
  {Pinsonneault}, {Gaidos}, {Garc{\'{\i}}a}, {Hekker}, {Mathur}, {Mosser},
  {Torres}, {Bastien}, {Basu}, {Bedding}, {Chaplin}, {Demory}, {Fleming},
  {Guo}, {Mann}, {Rowe}, {Serenelli}, {Smith}, \& {Stello}}]{Huber2014}
{Huber} D. {et~al.}, 2014, \apjs, 211, 2

\bibitem[{{Iglesias} \& {Rogers}(1996)}]{iglesias&rogers1996}
{Iglesias} C.~A., {Rogers} F.~J., 1996, \apj, 464, 943

\bibitem[{{Kurtz} {et~al}\mbox{.}(2014){Kurtz}, {Saio}, {Takata}, {Shibahashi},
  {Murphy}, \& {Sekii}}]{Kurtz2014}
{Kurtz} D.~W., {Saio} H., {Takata} M., {Shibahashi} H., {Murphy} S.~J., {Sekii}
  T., 2014, \mnras, 444, 102

\bibitem[{{Kurucz}(1993)}]{Kurucz1993}
{Kurucz} R., 1993, ATLAS9 Stellar Atmosphere Programs and 2 km/s grid.~Kurucz
  CD-ROM No.~13.~ Cambridge, Mass.: Smithsonian Astrophysical Observatory,
  1993., 13

\bibitem[{{Kurucz} \& {Bell}(1995)}]{Kurucz1995}
{Kurucz} R., {Bell} B., 1995, Atomic Line Data (R.L.~Kurucz and B.~Bell) Kurucz
  CD-ROM No.~23.~Cambridge, Mass.: Smithsonian Astrophysical Observatory,
  1995., 23

\bibitem[{{Kurucz}(2011)}]{Kurucz2011}
{Kurucz} R.~L., 2011, Canadian Journal of Physics, 89, 417

\bibitem[{{Landstreet} {et~al}\mbox{.}(2009){Landstreet}, {Kupka}, {Ford},
  {Officer}, {Sigut}, {Silaj}, {Strasser}, \& {Townshend}}]{Landstreet2009}
{Landstreet} J.~D., {Kupka} F., {Ford} H.~A., {Officer} T., {Sigut} T.~A.~A.,
  {Silaj} J., {Strasser} S., {Townshend} A., 2009, \aap, 503, 973

\bibitem[{{Lovisi} {et~al}\mbox{.}(2013{\natexlab{a}}){Lovisi}, {Mucciarelli},
  {Dalessandro}, {Ferraro}, \& {Lanzoni}}]{Lovisi2013}
{Lovisi} L., {Mucciarelli} A., {Dalessandro} E., {Ferraro} F.~R., {Lanzoni} B.,
  2013{\natexlab{a}}, \apj, 778, 64

\bibitem[{{Lovisi} {et~al}\mbox{.}(2010){Lovisi}, {Mucciarelli}, {Ferraro},
  {Lucatello}, {Lanzoni}, {Dalessandro}, {Beccari}, {Rood}, {Sills}, {Fusi
  Pecci}, {Gratton}, \& {Piotto}}]{Lovisi2010}
{Lovisi} L. {et~al.}, 2010, \apjl, 719, L121

\bibitem[{{Lovisi} {et~al}\mbox{.}(2013{\natexlab{b}}){Lovisi}, {Mucciarelli},
  {Lanzoni}, {Ferraro}, {Dalessandro}, \& {Monaco}}]{Lovisi2013a}
{Lovisi} L., {Mucciarelli} A., {Lanzoni} B., {Ferraro} F.~R., {Dalessandro} E.,
  {Monaco} L., 2013{\natexlab{b}}, \apj, 772, 148

\bibitem[{{Mathys}(1990)}]{Mathys1990}
{Mathys} G., 1990, \aap, 232, 151

\bibitem[{{McWilliam}(1998)}]{McWilliam1998}
{McWilliam} A., 1998, \aj, 115, 1640

\bibitem[{{Murphy} {et~al}\mbox{.}(2014){Murphy}, {Bedding}, {Shibahashi},
  {Kurtz}, \& {Kjeldsen}}]{murphyetal2014}
{Murphy} S.~J., {Bedding} T.~R., {Shibahashi} H., {Kurtz} D.~W., {Kjeldsen} H.,
  2014, \mnras, 441, 2515

\bibitem[{{Murphy} {et~al}\mbox{.}(2015){Murphy}, {Corbally}, {Gray}, {Cheng},
  {Neff}, {Koen}, {Kuehn}, {Newsome}, \& {Riggs}}]{murphy2015}
{Murphy} S.~J. {et~al.}, 2015, Pub. Ast. Soc. Australia, 32, e036

\bibitem[{{Murphy} {et~al}\mbox{.}(2016){Murphy}, {Fossati}, {Bedding}, {Saio},
  {Kurtz}, {Grassitelli}, \& {Wang}}]{Murphy2016}
{Murphy} S.~J., {Fossati} L., {Bedding} T.~R., {Saio} H., {Kurtz} D.~W.,
  {Grassitelli} L., {Wang} E.~S., 2016, \mnras, 459, 1201

\bibitem[{{Murphy} \& {Shibahashi}(2015)}]{murphy-shibahashi2015}
{Murphy} S.~J., {Shibahashi} H., 2015, \mnras, 450, 4475

\bibitem[{{Niemczura} {et~al}\mbox{.}(2015){Niemczura}, {Murphy}, {Smalley},
  {Uytterhoeven}, {Pigulski}, {Lehmann}, {Bowman}, {Catanzaro}, {van Aarle},
  {Bloemen}, {Briquet}, {De Cat}, {Drobek}, {Eyer}, {Gameiro}, {Gorlova},
  {Kami{\'n}ski}, {Lampens}, {Marcos-Arenal}, {P{\'a}pics}, {Vandenbussche},
  {Van Winckel}, {St{\c e}{\'s}licki}, \& {Fagas}}]{Niemczura2015}
{Niemczura} E. {et~al.}, 2015, \mnras, 450, 2764

\bibitem[{{Noguchi} {et~al}\mbox{.}(2002){Noguchi}, {Aoki}, {Kawanomoto},
  {Ando}, {Honda}, {Izumiura}, {Kambe}, {Okita}, {Sadakane}, {Sato}, {Tajitsu},
  {Takada-Hidai}, {Tanaka}, {Watanabe}, \& {Yoshida}}]{Noguchi2002}
{Noguchi} K. {et~al.}, 2002, \pasj, 54, 855

\bibitem[{{Paunzen} {et~al}\mbox{.}(1999){Paunzen}, {Andrievsky},
  {Chernyshova}, {Klochkova}, {Panchuk}, \& {Handler}}]{paunzen1999}
{Paunzen} E., {Andrievsky} S.~M., {Chernyshova} I.~V., {Klochkova} V.~G.,
  {Panchuk} V.~E., {Handler} G., 1999, \aap, 351, 981

\bibitem[{{Paxton} {et~al}\mbox{.}(2013){Paxton}, {Cantiello}, {Arras},
  {Bildsten}, {Brown}, {Dotter}, {Mankovich}, {Montgomery}, {Stello}, {Timmes},
  \& {Townsend}}]{Paxton2013}
{Paxton} B. {et~al.}, 2013, \apjs, 208, 4

\bibitem[{{Ryabchikova} {et~al}\mbox{.}(2004){Ryabchikova}, {Nesvacil},
  {Weiss}, {Kochukhov}, \& {St{\"u}tz}}]{Ryabchikova2004}
{Ryabchikova} T., {Nesvacil} N., {Weiss} W.~W., {Kochukhov} O., {St{\"u}tz} C.,
  2004, \aap, 423, 705

\bibitem[{{Saio} {et~al}\mbox{.}(2015){Saio}, {Kurtz}, {Takata}, {Shibahashi},
  {Murphy}, {Sekii}, \& {Bedding}}]{Saio2015}
{Saio} H., {Kurtz} D.~W., {Takata} M., {Shibahashi} H., {Murphy} S.~J., {Sekii}
  T., {Bedding} T.~R., 2015, \mnras, 447, 3264

\bibitem[{{Schmid} \& {Aerts}(2016)}]{Schmid2016}
{Schmid} V.~S., {Aerts} C., 2016, \aap, 592, A116

\bibitem[{{Shibahashi} \& {Kurtz}(2012)}]{Shibahashi2012}
{Shibahashi} H., {Kurtz} D.~W., 2012, \mnras, 422, 738

\bibitem[{{Sills}, {Adams} \& {Davies}(2005){Sills}, {Adams}, \&
  {Davies}}]{sills2005}
{Sills} A., {Adams} T., {Davies} M.~B., 2005, \mnras, 358, 716

\bibitem[{{Smith}(1971)}]{Smith1971}
{Smith} M.~A., 1971, \aap, 11, 325

\bibitem[{{St{\c e}pie{\'n}}(2000)}]{Stepien2000}
{St{\c e}pie{\'n}} K., 2000, \aap, 353, 227

\bibitem[{{St\"urenburg}(1993)}]{Stuerenburg1993}
{St\"urenburg} S., 1993, \aap, 277, 139

\bibitem[{{Takeda} {et~al}\mbox{.}(2012){Takeda}, {Kang}, {Han}, {Lee}, {Kim},
  {Kawanomoto}, \& {Ohishi}}]{Takeda2012}
{Takeda} Y., {Kang} D., {Han} I., {Lee} B., {Kim} K., {Kawanomoto} S., {Ohishi}
  N., 2012, \pasj, 64, 38

\bibitem[{{Takeda}, {Ohkubo} \& {Sadakane}(2002){Takeda}, {Ohkubo}, \&
  {Sadakane}}]{Takeda2002}
{Takeda} Y., {Ohkubo} M., {Sadakane} K., 2002, \pasj, 54, 451

\bibitem[{{Takeda} {et~al}\mbox{.}(2005){Takeda}, {Ohkubo}, {Sato}, {Kambe}, \&
  {Sadakane}}]{Takeda2005}
{Takeda} Y., {Ohkubo} M., {Sato} B., {Kambe} E., {Sadakane} K., 2005, \pasj,
  57, 27

\bibitem[{{Takeda}, {Sato} \& {Murata}(2008){Takeda}, {Sato}, \&
  {Murata}}]{Takeda2008b}
{Takeda} Y., {Sato} B., {Murata} D., 2008, \pasj, 60, 781

\bibitem[{{Westin} {et~al}\mbox{.}(2000){Westin}, {Sneden}, {Gustafsson}, \&
  {Cowan}}]{Westin2000}
{Westin} J., {Sneden} C., {Gustafsson} B., {Cowan} J.~J., 2000, \apj, 530, 783

\end{thebibliography}

\appendix

\section{Atomic line data for F{\MakeLowercase e} lines}
\label{app:fe}

\begin{table}
\caption{Data for 67 Fe~{\sc i} lines selected for analysis. $\chi$ is the lower excitation potential.}
\scriptsize
\centering
\begin{tabular}{ccrr}
\toprule
  \multicolumn{1}{c}{Wavelength}  &  \multicolumn{1}{c}{$\chi$}  &  \multicolumn{1}{c}{log $gf$} &   \multicolumn{1}{c}{$W_{\lambda}$} \\
     \multicolumn{1}{c}{(\AA)}    &    \multicolumn{1}{c}{(eV)}   &         &   \multicolumn{1}{c}{(m\AA)}      \\
\hline
  
 4430.614 & 2.223 & $-$1.728  &   40.0    \\
 4442.338 & 2.198 & $-$1.228  &   65.0    \\
 4447.718 & 2.223 & $-$1.339  &  56.4  \\
 4489.739 & 0.121 & $-$3.899  &     6.0 \\
 4494.563 & 2.198 & $-$1.143  &  77.1 \\
  4602.940 & 1.485 & $-$2.208  &  30.3 \\
 4736.772 & 3.211 & $-$0.752  &  41.1 \\
 4872.136 & 2.882 & $-$0.567  &  60.6 \\
 4890.754 & 2.875 & $-$0.394  &  81.2 \\
 4891.492 & 2.851 & $-$0.111  & 103.2 \\
 4903.308 & 2.882 & $-$0.926  &  48.6 \\
 4918.993 & 2.865 & $-$0.342  &   79.0 \\
 4938.813 & 2.875 & $-$1.077  &  30.4 \\
 4946.385 & 3.368 &  $-$1.010  &  12.7 \\
 4966.087 & 3.332 &  $-$0.840  &  32.4 \\
 4973.101 &  3.960 &  $-$0.850  &  18.8 \\
 4994.129 & 0.915 & $-$2.969  &  15.8 \\
 5014.941 & 3.943 &  $-$0.270  &   43.0 \\
 5022.236 & 3.984 &  $-$0.490  &  23.6 \\
 5049.819 & 2.279 & $-$1.355  &  49.9 \\
 5068.765 &  2.940 & $-$1.041  &  33.5 \\
 5074.748 &  4.220 &  $-$0.160  &  41.9 \\
 5083.338 & 0.958 & $-$2.842  &  16.5 \\
 5090.767 & 4.256 &  $-$0.360  &  17.9 \\
 5123.719 & 1.011 & $-$3.057  &  14.1 \\
 5133.681 & 4.178 &     0.200 &  60.2 \\
 5162.292 & 4.178 &   0.020  &  59.3 \\
 5192.343 & 2.998 & $-$0.421  &   73.0 \\
 5198.711 & 2.223 &  $-$2.140  &  15.5 \\                           
 5216.274 & 1.608 & $-$2.082  &  31.8  \\
 5232.939 &  2.940 & $-$0.057  & 101.4 \\
 5242.491 & 3.634 &  $-$0.970  &  17.2 \\
 5266.555 & 2.998 & $-$0.385  &  77.2 \\
  5281.790 & 3.038 & $-$0.833  &  43.8 \\
 5283.621 & 3.241 & $-$0.524  &  56.4 \\
 5339.928 & 3.266 &  $-$0.720  &   40.0 \\
 5364.858 & 4.446 &   0.230  &  46.2 \\
 5367.479 & 4.415 &   0.440  &  55.2 \\
 5369.958 & 4.371 &  0.536  &  64.4 \\
 5383.369 & 4.312 &  0.645  &  76.9 \\
 5389.479 & 4.415 &  $-$0.410  &  17.1 \\
 5393.167 & 3.241 & $-$0.715  &  38.7 \\
 5400.502 & 4.371 &   $-$0.100  &  32.3 \\
 5405.774 &  0.990 & $-$1.852  &  84.9 \\
 5410.910 & 4.473 & 0.400 &  49.4    \\
 5424.069 &  4.320 &   0.520  &  83.8 \\ 
 5434.523 & 1.011 & $-$2.126  &  66.6 \\
 5445.042 & 4.386 &   0.040  &  39.8 \\
 5446.916 &  0.990 &  $-$1.910  &  94.3 \\
 5569.618 & 3.417 &   $-$0.500  &  44.4 \\
 5572.841 & 3.396 & $-$0.275  &  59.6 \\
  5576.090 &  3.430 &   $-$0.900  &  23.5 \\
 5586.756 & 3.368 & $-$0.096  &  72.7 \\
  6020.170 & 4.607 &  $-$0.210  &  17.7 \\
 6024.049 & 4.548 &  $-$0.060  &  30.9 \\
 6136.615 & 2.453 &  $-$1.410  &  45.6 \\
 6137.694 & 2.588 & $-$1.346  &  36.8 \\
 6213.429 & 2.223 & $-$2.481  &   3.7 \\
 6219.279 & 2.198 & $-$2.448  &     6.0 \\
 6230.726 & 2.559 & $-$1.276 &  51.0 \\
 6252.554 & 2.404 & $-$1.767 & 30.5 \\
 6265.131 & 2.176 &  $-$2.550 &  7.2 \\
 6301.498 & 3.654 & $-$0.718 & 35.1 \\
 6393.602 & 2.433 & $-$1.576 & 28.7 \\
 6400.000 & 3.602 &  $-$0.290 & 53.3 \\
 6421.349 & 2.279 & $-$2.014 & 13.2 \\
 6592.913 & 2.727 &  $-$1.470 & 21.5 \\
\bottomrule
\end{tabular} 
 \label{tab:02}
\end{table}

\begin{table}
\caption{Data for 21 Fe~{\sc ii} lines selected for analysis. 
}
\centering
\begin{tabular}{cccrr}
\toprule
  \multicolumn{1}{c}{Wavelength}  &  \multicolumn{1}{c}{$\chi$}  &  \multicolumn{1}{c}{log $gf$} &   \multicolumn{1}{c}{$W_{\lambda}$} \\
   \multicolumn{1}{c}{(\AA)}    &    \multicolumn{1}{c}{(eV)}   &         &   \multicolumn{1}{c}{(m\AA)}      \\
\hline

   4416.830 &     2.778 &      $-$2.580 &      96.9 \\
  4491.405 &     2.855 &      $-$2.590 &      69.6 \\
  4508.288 &     2.855 &     $-$2.318 &     110.3 \\
  4515.339 &     2.844 &     $-$2.422 &      88.2 \\
  4520.224 &     2.807 &      $-$2.590 &      86.4 \\
  4541.524 &     2.855 &      $-$3.030 &       74.0 \\
   4576.340 &     2.844 &      $-$2.920 &      63.1 \\
  4620.521 &     2.828 &     $-$3.226 &      30.5 \\
  4993.358 &     2.807 &      $-$3.650 &      15.8 \\
  5197.577 &      3.230 &     $-$2.167 &       90.0 \\
  5234.625 &     3.221 &     $-$2.268 &      94.7 \\
  5264.812 &      3.230 &     $-$3.108 &      29.7 \\
  5276.002 &     3.199 &     $-$1.963 &       85.0 \\
  5284.109 &     2.891 &      $-$3.010 &      42.2 \\ 
  5325.553 &     3.221 &      $-$3.210 &      22.3 \\
  5425.257 &     3.199 &      $-$3.210 &      18.6 \\
  5534.847 &     3.245 &      $-$2.770 &      49.3 \\
  6084.111 &     3.199 &     $-$3.791 &       6.4 \\
  6149.258 &     3.889 &      $-$2.720 &      17.5 \\
  6247.557 &     3.892 &      $-$2.340 &      39.6 \\
   6432.680 &     2.891 &      $-$3.580 &      17.7 \\

\bottomrule
\end{tabular} 
 \label{tab:03}
\end{table}

\clearpage
\section{Data for each element and line}

\begin{table*}
\caption{
Atomic data, $W_{\lambda}$,  and the abundances  derived from each line of detected  elements together with the number ($N$) of lines  and  average abundance with standard deviation.}
\scriptsize
\centering
\begin{tabular}{rccrrrr}
 \toprule
  Ion/Code   &  Wavelength  & $\chi$  &  log $gf$  &  $W_{\lambda}$ &  log X &  Remark \\
             &    (\AA)       &  (eV)    &      &      (m\AA)   &          &   \\
 \hline
             &          &        &           &        &      &     \\
Li~{\sc i}   &   $N=1$   &      &      &      &  $\le 1.38$    &     \\
 \hline 
3.00   &   6707.773   &   0.000   &   0.002   & $\le  0.4$   &  $ \le 1.38$   &
 single line \\
   &      &      &      &      &      &          \\
C~{\sc i}   &   $N=10$  &      &      &      &   7.82$\pm 0.15$   &      \\
 \hline  
6.00   &   4770.027   &   7.483   &   $-$2.439   &   8.3   &   8.03   &    \\
6.00   &   4932.049   &   7.685   &   $-$1.658   &   24.4   &   7.91   &   \\
6.00   &   5052.167   &   7.685   &   $-$1.304   &   42.2   &   7.85   &    \\
6.00   &   5380.337   &   7.685   &   $-$1.615   &   24.4   &   7.87   &    \\
6.00   &   6013.166   &   8.647   &   $-$1.370   &   11.8   &   7.60   &
 doublet fit  \\
6.00   &   6014.830   &   8.643   &   $-$1.710   &   4.4   &   7.84   &    \\
6.00   &   7111.472   &   8.640   &   $-$1.086   &   8.5   &   7.54   &    \\
6.00   &   7113.178   &   8.647   &   $-$0.774   &   20.7   &   7.66   &    \\
6.00   &   7115.172   &   8.643   &   $-$0.935   &   26.9   &   7.87   &
 doublet fit \\
6.00   &   7116.991   &   8.647   &   $-$0.907   &   29.5   &   7.98   &    \\
   &      &      &      &      &      &     \\
O~{\sc i}   &   $N=3$   &      &      &      &   8.38$\pm 0.09$   &    \\
  \hline  
8.00   &   6155.961   &   10.740   &   $-$1.363   &   6.9   &   8.51 & triplet fit \\
8.00   &   6156.737   &   10.740   &   $-$1.488   &   11.7   &   8.31   &
 triplet fit \\
8.00   &   6158.149   &   10.741   &   $-$1.841   &   14.3   &   8.32   &
 triplet fit\\
   &      &      &      &      &      &    \\
Na~{\sc i}   &   $N=4$   &      &      &      &   5.75$\pm 0.04$   &     \\
 \hline 
11.00   &   5682.633   &   2.102   &   $-$0.700   &   20.8   &   5.81   &    \\
11.00   &   5688.194   &   2.104   &   $-$1.400   &   31.4   &   5.73   &   doublet fit\\
11.00   &   6154.226   &   2.102   &   $-$1.560   &   2.8   &   5.74   &     \\
11.00   &   6160.747   &   2.104   &   $-$1.260   &   5.2   &   5.71   &     \\
   &      &      &      &      &      &      \\
Mg~{\sc i}   &   $N=3$   &      &      &      &   7.13$\pm 0.04$   &    \\
 \hline
12.00   &   4571.096   &   0.000   &   $-$5.691   &   5.7   &   7.10   &      \\
12.00   &   4730.029   &   4.346   &   $-$2.523   &   6.2   &   7.19   &     \\
12.00   &   5711.088   &   4.346   &   $-$1.833   &   23.0   &   7.11   &     \\
   &      &      &      &      &      &        \\
Mg~{\sc ii}   &   $N=1$  &      &      &      &   7.29   &      \\
 \hline 
12.01   &   4481.126   &   8.863   &   0.730   &   292.7   &   7.29   &   triplet fit   \\
   &      &      &      &      &      &   \\
Si~{\sc i}   &   $N=14$   &      &      &      &   7.05$\pm 0.22$   &   \\
 \hline 
14.00   &   5665.555   &   4.920   &   $-$2.040   &   4.0   &   7.01   &    \\
14.00   &   5675.417   &   5.619   &   $-$1.030   &   10.5   &   6.94   &   \\
14.00   &   5675.756   &   5.619   &   $-$1.780   &   4.2   &   7.28   &  \\
14.00   &   5684.484   &   4.954   &   $-$1.650   &   15.3   &   7.26   &    \\
14.00   &   5690.425   &   4.930   &   $-$1.870   &   8.0   &   7.16   &     \\
14.00   &   5948.541   &   5.082   &   $-$1.230   &   24.0   &   7.16   &     \\
14.00   &   6091.919   &   5.871   &   $-$1.400   &   4.9   &   7.15   &    \\ 
14.00   &   6155.134   &   5.619   &   $-$0.400   &   25.7   &   6.76   &   \\ 
14.00   &   6243.815   &   5.616   &   $-$0.770   &   11.1   &   6.71   &    \\
14.00   &   6244.466   &   5.616   &   $-$0.690   &   11.8   &   6.66   &    \\
14.00   &   6254.188   &   5.619   &   $-$0.600   &   25.3   &   6.95   &    \\
14.00   &   6721.848   &   5.862   &   $-$1.490   &   7.2   &   7.41   &     \\
14.00   &   7003.569   &   5.964   &   $-$0.970   &   12.1   &   7.19   &     \\
14.00   &   7034.901   &   5.871   &   $-$0.880   &   13.8   &   7.10   &      \\
  &      &      &      &      &      &   \\
Si~{\sc ii}   &   $N=3$   &      &      &      &   7.03$\pm 0.07$   &     \\
 \hline
14.01   &   5957.559   &   10.066   &   $-$0.349   &   3.2   &   6.93   &   \\  
14.01   &   5978.930   &   10.074   &   $-$0.061   &   8.4   &   7.10   &     \\
14.01   &   6347.109   &   8.121   &   0.297   &   104.3   &   7.07   &    \\
 
\bottomrule
\end{tabular}
\label{tab:A}
\end{table*}


\begin{table*}
\contcaption{}
\scriptsize
\centering
\begin{tabular}{rccrrrr}
 \toprule

  Ion/Code   &  Wavelength  & $\chi$  &  log $gf$  &  $W_{\lambda}$ &  log X &  Remark \\
             &    (\AA)       &  (eV)    &      &      (m\AA)   &          &   \\
 \hline
             &          &        &           &        &      &     \\

S~{\sc i}   &   $N=8$   &      &      &      &   6.87$\pm 0.08$   &     \\
 \hline
16.00   &   4694.113   &   6.524   &   $-$1.770   &   9.1   &   6.91   &     \\
16.00   &   4695.443   &   6.524   &   $-$1.920   &   6.1   &   6.88   &     \\
16.00   &   4696.252   &   6.524   &   $-$2.140   &   5.0   &   7.01   &     \\
16.00   &   6045.954   &   7.867   &   $-$1.820   &   7.1   &   6.83   &   triplet fit  \\
16.00   &   6052.583   &   7.870   &   $-$1.330   &   9.2   &   6.77   &   doublet fit  \\
16.00   &   6743.440   &   7.866   &   $-$1.270   &   10.0   &   6.74   &   triplet fit   \\
16.00   &   6748.573   &   7.867   &   $-$1.390   &   20.0   &   6.95   &  triplet fit   \\
16.00   &   6756.851   &   7.870   &   $-$1.760   &   26.1   &   6.85   &   triplet fit  \\
   &      &      &      &      &      &     \\
Ca~{\sc i}   &   $N=25$   &      &      &      &   5.95$\pm 0.12$   &   \\
 \hline 
20.00   &   4425.437   &   1.879   &   $-$0.385   &   82.2   &   6.01   &    \\
20.00   &   4434.957   &   1.886   &   $-$0.029   &   116.5   &   6.09   &    \\
20.00   &   4435.679   &   1.886   &   $-$0.500   &   77.0   &   6.07   &     \\
20.00   &   4454.779   &   1.899   &   0.252   &   132.8   &   6.05   &    \\
20.00   &   4455.887   &   1.899   &   $-$0.510   &   64.3   &   5.93   &     \\
20.00   &   4526.928   &   2.709   &   $-$0.430   &   25.9   &   5.87   &    \\
20.00   &   5261.704   &   2.521   &   $-$0.730   &   28.8   &   6.07   &    \\
20.00   &   5265.556   &   2.523   &   $-$0.260   &   55.2   &   6.00   &     \\
20.00   &   5581.965   &   2.523   &   $-$0.710   &   32.8   &   6.12   &    \\
20.00   &   5588.749   &   2.526   &   0.210   &   93.1   &   6.01   &    \\
20.00   &   5590.114   &   2.521   &   $-$0.710   &   30.7   &   6.08   &    \\
20.00   &   5594.462   &   2.523   &   $-$0.050   &   74.2   &   6.03   &    \\
20.00   &   5601.277   &   2.526   &   $-$0.690   &   35.8   &   6.15   &    \\
20.00   &   5857.451   &   2.932   &   0.230   &   65.5   &   5.94   &  \\
20.00   &   6102.723   &   1.879   &   $-$0.890   &   46.4   &   6.01   &    \\
20.00   &   6122.217   &   1.886   &   $-$0.409   &   90.2   &   6.08   &    \\
20.00   &   6162.173   &   1.899   &   0.100   &   105.9   &   5.77   &    \\
20.00   &   6166.439   &   2.521   &   $-$0.900   &   11.2   &   5.74   &    \\
20.00   &   6169.042   &   2.523   &   $-$0.550   &   24.7   &   5.79   &    \\
20.00   &   6169.563   &   2.526   &   $-$0.270   &   37.3   &   5.74   &    \\
20.00   &   6439.075   &   2.526   &   0.470   &   105.0   &   5.88   &    \\
20.00   &   6462.567   &   2.523   &   0.310   &   88.1   &   5.82   &    \\
20.00   &   6471.662   &   2.526   &   $-$0.590   &   23.8   &   5.81   &    \\
20.00   &   6493.781   &   2.521   &   0.140   &   75.4   &   5.84   &    \\
20.00   &   6717.681   &   2.709   &   $-$0.610   &   24.4   &   5.97   &    \\
   &      &      &      &      &      &    \\
Ca~{\sc ii}   &   $N=3$   &      &      &      &   6.02$\pm 0.09$   &    \\
 \hline
20.01   &   5001.479   &   7.505   &   $-$0.517   &   20.7   &   5.91   &   \\  
20.01   &   5019.971   &   7.515   &   $-$0.257   &   40.4   &   6.02   &    \\
20.01   &   5285.266   &   7.505   &   $-$1.153   &   8.6   &   6.13   &     \\
   &      &      &      &      &      &     \\
Sc~{\sc ii}   &   $N=9 $  &      &      &      &   2.8$\pm 0.15$   &    \\
 \hline
21.01   &   4400.389   &   0.605   &   $-$0.510   &   82.8   &   2.84   & hfs fit \\
21.01   &   4415.557   &   0.595   &   $-$0.640   &   98.5   &   2.97   & hfs fit \\
21.01   &   5031.021   &   1.357   &   $-$0.260   &   45.9   &   2.58   & hfs fit \\
21.01   &   5526.790   &   1.768   &    0.130   &   66.9   &   2.75   & hfs fit \\
21.01   &   5641.001   &   1.500   &   $-$1.040   &   14.0   &   2.80   &    \\
21.01   &   5667.149   &   1.500   &   $-$1.240   &   10.1   &   2.85   &    \\
21.01   &   5684.202   &   1.507   &   $-$1.050   &   14.6   &   2.84   &    \\
21.01   &   6245.637   &   1.507   &   $-$0.980   &   9.1   &   2.52   &    \\
21.01   &   6604.601   &   1.357   &   $-$1.480   &   11.7   &   3.02   &    \\
   &      &      &      &      &      &    \\
\bottomrule
\end{tabular}
\label{tab:A1}
\end{table*}

\begin{table*}
\contcaption{}
\scriptsize
\centering
\begin{tabular}{rccrrrr}
 \toprule

  Ion/Code   &  Wavelength  & $\chi$  &  log $gf$  &  $W_{\lambda}$ &  log X &  Remark \\
             &    (\AA)       &  (eV)    &      &      (m\AA)   &          &   \\
 \hline
             &          &        &           &        &      &     \\
Ti~{\sc i}   &   $N=15$   &      &      &      &   4.59$\pm 0.11$   &   \\
 \hline 
22.00   &   4449.143   &   1.887   &   0.500   &   9.5   &   4.69   &    \\
22.00   &   4518.023   &   0.826   &   $-$0.325   &   5.7   &   4.49   &   \\
22.00   &   4534.778   &   0.836   &   0.280   &   24.5   &   4.58   &   \\
22.00   &   4548.765   &   0.826   &   $-$0.354   &   7.7   &   4.65   &   \\
22.00   &   4681.908   &   0.048   &   $-$1.071   &   6.1   &   4.66   &   \\
22.00   &   4759.272   &   2.256   &   0.514   &   3.9   &   4.53   &   \\
22.00   &   4981.732   &   0.848   &   0.504   &   34.8   &   4.54   &   \\
22.00   &   4999.504   &   0.826   &   0.250   &   27.5   &   4.64   &   \\
22.00   &   5022.871   &   0.826   &   $-$0.434   &   3.8   &   4.38   &   \\
22.00   &   5036.468   &   1.443   &   0.130   &   10.9   &   4.76   &   \\
22.00   &   5173.742   &   0.000   &   $-$1.118   &   4.6   &   4.52   &   \\
22.00   &   5192.969   &   0.021   &   $-$1.006   &   7.0   &   4.61   &   \\
22.00   &   5210.386   &   0.048   &   $-$0.884   &   7.0  &   4.51   &   \\
22.00   &   6258.104   &   1.443   &   $-$0.355   &   3.2   &   4.65   &   \\
22.00   &   6258.709   &   1.460   &   $-$0.240   &   6.3   &   4.85   &   \\
   &      &      &      &      &      &    \\
Ti~{\sc ii}    &   $N=32$   &      &      &      &   4.59$\pm 0.18$   &   \\
 \hline 
22.01   &   4409.243   &   1.243   &   $-$2.638   &   15.8   &   4.71   &  \\
22.01   &   4409.516   &   1.231   &   $-$2.569   &   15.1   &   4.60   &   \\
22.01   &   4411.925   &   1.224   &   $-$2.520   &   10.5   &   4.38   &   \\
22.01   &   4421.938   &   2.061   &   $-$1.770   &   29.9   &   4.78   &   \\
22.01   &   4444.558   &   1.116   &   $-$2.030   &   40.8   &   4.50   &   \\
22.01   &   4450.482   &   1.084   &   $-$1.450   &   104.1   &   4.69   &  \\
22.01   &   4464.450   &   1.161   &   $-$2.080   &   72.8   &   4.99   &   \\
22.01   &   4470.857   &   1.165   &   $-$2.280   &   37.8   &   4.74   &  \\
22.01   &   4488.331   &   3.123   &   $-$0.820   &   51.8   &   4.96   &  \\
22.01   &   4441.734   &   1.180   &   $-$2.410   &   30.3   &   4.76   &   \\
22.01   &   4518.327   &   1.080   &   $-$2.555   &   14.9   &   4.46   &   \\
22.01   &   4529.474   &   1.572   &   $-$2.030   &   50.7   &   4.98   &   \\
22.01   &   4544.028   &   1.243   &   $-$2.400   &   16.8   &   4.48   &   \\
22.01   &   4568.314   &   1.224   &   $-$2.650   &   7.1   &   4.31   &   \\
22.01   &   4708.665   &   1.237   &   $-$2.340   &   21.6   &   4.53   &  \\
22.01   &   4764.526   &   1.237   &   $-$2.950   &   10.4   &   4.79   &  \\
22.01   &   4779.985   &   2.048   &   $-$1.370   &   49.3   &   4.65   &  \\
22.01   &   4798.521   &   1.080   &   $-$2.430   &   15.5   &   4.34   &   \\
22.01   &   4805.085   &   2.061   &   $-$1.100   &   77.4   &   4.74   &   \\
22.01   &   4911.193   &   3.123   &   $-$0.340   &   36.3   &   4.24   &  \\
22.01   &   5010.212   &   3.095   &   $-$1.290   &   12.3   &   4.60   &   \\
22.01   &   5013.677   &   1.582   &   $-$1.935   &   24.0   &   4.43   &   \\
22.01   &   5129.152   &   1.892   &   $-$1.390   &   62.4   &   4.69   &  \\
22.01   &   5154.070   &   1.566   &   $-$1.750   &   42.0   &   4.53   &   \\
22.01   &   5185.913   &   1.893   &   $-$1.350   &   50.9   &   4.50   &   \\
22.01   &   5211.536   &   2.590   &   $-$1.356   &   16.1   &   4.40   &   \\
22.01   &   5336.786   &   1.582   &   $-$1.590   &   56.3   &   4.57   &  \\
22.01   &   5381.021   &   1.566   &   $-$1.920   &   30.5   &   4.51   &  \\
22.01   &   5418.764   &   1.582   &   $-$2.000   &   24.6   &   4.48   &   \\
22.01   &   5490.690   &   1.566   &   $-$2.650   &   7.3   &   4.53   &  \\
22.01   &   6491.561   &   2.061   &   $-$1.793   &   18.1   &   4.45   &   \\  
22.01   &   6998.905   &   3.123   &   $-$1.453   &   11.3   &   4.68   &   \\
   &      &      &      &      &      &   \\
Cr~{\sc i}   &  $N=14$   &      &      &      &   4.9$\pm 0.11$   &   \\
 \hline 
24.00   &   4545.953   &   0.941   &   $-$1.370   &   8.6   &   5.07   &   \\
24.00   &   4616.120   &   0.983   &   $-$1.190   &   4.9   &   4.66   &    \\
24.00   &   4626.174   &   0.968   &   $-$1.320   &   5.8   &   4.85   &    \\
24.00   &   4646.162   &   1.030   &   $-$0.700   &   21.6   &   4.9   &      \\
24.00   &   4651.282   &   0.983   &   $-$1.460   &   3.8   &   4.81   &    \\
24.00   &   4652.152   &   1.004   &   $-$1.030   &   9.4   &   4.81   &    \\
24.00   &   4718.426   &   3.195   &   0.090   &   5.8   &   5.09   &     \\
24.00   &   5204.506   &   0.941   &   $-$0.208   &   62.0   &   4.95   &    \\
24.00   &   5206.038   &   0.941   &   0.019   &   80.3   &   4.95   &   \\
24.00   &   5296.691   &   0.983   &   $-$1.400   &   6.0   &   4.92   &    \\
24.00   &   5297.376   &   2.899   &   0.167   &   9.0   &   4.97   &    \\
24.00   &   5298.277   &   0.983   &   $-$1.150   &   8.9   &   4.85   &    \\
24.00   &   5348.312   &   1.004   &   $-$1.290   &   8.9   &   5.01   &    \\
24.00   &   6978.488   &   3.463   &   0.142   &   2.5   &   4.81   &    \\

\bottomrule
\end{tabular}
\label{tab:A2}
\end{table*}


\begin{table*}
\contcaption{}
\scriptsize
\centering
\begin{tabular}{rccrrrr}
 \toprule

  Ion/Code   &  Wavelength  & $\chi$  &  log $gf$  &  $W_{\lambda}$ &  log X &  Remark \\
             &    (\AA)       &  (eV)    &      &      (m\AA)   &          &   \\
 \hline
             &          &        &           &        &      &     \\
Cr~{\sc ii}   &  $ N=21$   &      &      &      &   4.96$\pm 0.12$   &   \\
 \hline 
24.01   &   4558.650   &   4.073   &   $-$0.410   &   68.9   &   4.70   &    \\
24.01   &   4592.049   &   4.073   &   $-$1.217   &   22.3   &   4.79   &    \\
24.01   &   4616.629   &   4.072   &   $-$1.291   &   22.3   &   4.86   &    \\
24.01   &   4618.803   &   4.073   &   $-$0.860   &   56.6   &   4.99   &    \\
24.01   &   4634.070   &   4.072   &   $-$0.990   &   42.5   &   4.92   &     \\
24.01   &   4812.337   &   3.864   &   $-$1.995   &   11.5   &   5.09   &    \\
24.01   &   4824.127   &   3.871   &   $-$0.970   &   67.0   &   5.10   &     \\
24.01   &   4836.229   &   3.858   &   $-$2.000   &   11.5   &   5.12   &     \\
24.01   &   4848.235   &   3.864   &   $-$1.140   &   43.3   &   5.06   &    \\
24.01   &   4884.607   &   3.858   &   $-$2.080   &   6.3   &   4.93   &     \\
24.01   &   5237.329   &   4.073   &   $-$1.160   &   40.4   &   5.04   &    \\
24.01   &   5246.768   &   3.713   &   $-$2.450   &   3.2   &   4.82   &    \\
24.01   &   5305.853   &   3.827   &   $-$2.080   &   7.9   &   4.94   &    \\
24.01   &   5308.440   &   4.071   &   $-$1.810   &   6.1   &   4.73   &    \\
24.01   &   5313.590   &   4.073   &   $-$1.650   &   16.9   &   5.06   &    \\
24.01   &   5334.869   &   4.072   &   $-$1.562   &   17.7   &   4.99   &    \\
24.01   &   5407.604   &   3.827   &   $-$2.088   &   7.6   &   4.93   &    \\
24.01   &   5420.922   &   3.758   &   $-$2.360   &   5.0   &   4.96   &     \\
24.01   &   5502.067   &   4.168   &   $-$1.990   &   7.1   &   5.05   &    \\
24.01   &   5508.606   &   4.156   &   $-$2.110   &   6.6   &   5.13   &    \\
24.01   &   5510.702   &   3.827   &   $-$2.452   &   3.6   &   4.95   &   \\
   &      &      &      &      &      &    \\
Mn~{\sc i}  &  $ N=7$   &      &      &      &   4.57$\pm 0.05$   &   \\
 \hline 
25.00   &   4709.712   &   2.888   &   $-$0.340   &   4.5   &   4.64   &    \\
25.00   &   4754.042   &   2.282   &   $-$0.086   &   16.4   &   4.53   &    \\
25.00   &   4765.846   &   2.941   &   $-$0.080   &   7.0   &   4.61   &    \\
25.00   &   4766.418   &   2.920   &   0.100   &   10.8   &   4.62   &    \\
25.00   &   4783.427   &   2.298   &   0.042   &   24.0   &   4.58   & hfs fit \\
25.00   &   4823.524   &   2.319   &   0.144   &   26.6   &   4.50   &
 hfs fit \\
25.00   &   6021.79   &   3.075   &   0.034   &   6.6   &   4.53   &    \\
   &      &      &      &      &      &    \\
Ni~{\sc i}    &  $ N=41$   &      &      &      &   5.7$\pm 0.10$   &   \\
 \hline 
28.00   &   4648.646   &   3.419   &   $-$0.100   &   19.9   &   5.52   &    \\
28.00   &   4686.207   &   3.597   &   $-$0.580   &   7.1   &   5.63   &    \\
28.00   &   4703.803   &   3.658   &   $-$0.735   &   6.1   &   5.76   &    \\
28.00   &   4714.408   &   3.380   &   0.260   &   52.5   &   5.69   &     \\
28.00   &   4715.757   &   3.543   &   $-$0.320   &   12.8   &   5.61   &    \\
28.00   &   4732.456   &   4.105   &   $-$0.550   &   4.8   &   5.80   &    \\
28.00   &   4756.510   &   3.480   &   $-$0.270   &   17.0   &   5.65   &    \\
28.00   &   4786.531   &   3.419   &   $-$0.160   &   25.6   &   5.71   &    \\
28.00   &   4829.016   &   3.542   &   $-$0.330   &   13.6   &   5.67   &    \\
28.00   &   4873.438   &   3.699   &   $-$0.380   &   7.3   &   5.65   &     \\
28.00   &   4904.407   &   3.542   &   $-$0.170   &   21.3   &   5.72   &    \\
28.00   &   4918.362   &   3.841   &   $-$0.240   &   11.6   &   5.70   &    \\
28.00   &   4925.559   &   3.655   &   $-$0.770   &   5.1   &   5.71   &    \\
28.00   &   4935.831   &   3.941   &   $-$0.350   &   11.2   &   5.86   &    \\
28.00   &   4937.341   &   3.606   &   $-$0.390   &   9.6   &   5.58   &    \\
28.00   &   4953.200   &   3.740   &   $-$0.580   &   5.6   &   5.62   &    \\
28.00   &   4965.166   &   3.796   &   $-$0.753   &   7.0   &   5.94   &    \\
28.00   &   4980.166   &   3.606   &   0.070   &   26.8   &   5.63   &    \\
28.00   &   5000.338   &   3.635   &   $-$0.430   &   14.0   &   5.82   &    \\
28.00   &   5010.934   &   3.635   &   $-$0.870   &   3.7   &   5.65   &    \\
28.00   &   5017.568   &   3.539   &   $-$0.020   &   24.4   &   5.62   &    \\
28.00   &   5035.357   &   3.635   &   0.290   &   37.2   &   5.61   &    \\
28.00   &   5042.182   &   3.658   &   $-$0.580   &   7.2   &   5.67   &    \\
28.00   &   5048.843   &   3.847   &   $-$0.380   &   6.2   &   5.54   &   \\
28.00   &   5080.528   &   3.655   &   0.330   &   42.8   &   5.67   &    \\
28.00   &   5081.107   &   3.847   &   0.300   &   32.4   &   5.68   &    \\
28.00   &   5082.339   &   3.658   &   $-$0.540   &   7.3   &   5.64   &    \\
28.00   &   5084.089   &   3.678   &   0.030   &   26.1   &   5.69   &    \\
28.00   &   5099.927   &   3.678   &   $-$0.100   &   29.1   &   5.89   &    \\
28.00   &   5115.389   &   3.834   &   $-$0.110   &   18.1   &   5.77   &    \\
28.00   &   5146.480   &   3.706   &   0.120   &   22.2   &   5.55   &     \\
28.00   &   5155.762   &   3.898   &   $-$0.090   &   14.5   &   5.68   &    \\
28.00   &   5176.559   &   3.898   &   $-$0.440   &   5.4   &   5.57   &    \\

\bottomrule
\end{tabular}
\label{tab:A3}
\end{table*}


\begin{table*}
\contcaption{}
\scriptsize
\centering
\begin{tabular}{rccrrrr}
 \toprule

  Ion/Code   &  Wavelength  & $\chi$  &  log $gf$  &  $W_{\lambda}$ &  log X &  Remark \\
             &    (\AA)       &  (eV)    &      &      (m\AA)   &          &   \\
 \hline
             &          &        &           &        &      &     \\
28.00   &   5614.768   &   4.153   &   $-$0.508   &   4.3   &   5.71   &    \\
28.00   &   5682.198   &   4.105   &   $-$0.470   &   7.2   &   5.87   &    \\
28.00   &   6086.276   &   4.266   &   $-$0.530   &   4.1   &   5.79   &    \\
28.00   &   6176.807   &   4.088   &   $-$0.260   &   7.3   &   5.65   &    \\
28.00   &   6191.171   &   1.676   &   $-$2.353   &   4.4   &   5.69   &    \\
28.00   &   6643.629   &   1.676   &   $-$2.300   &   4.8   &   5.67   &     \\
28.00   &   6767.768   &   1.826   &   $-$2.170   &   7.3   &   5.84   &     \\
28.00   &   6772.313   &   3.658   &   $-$0.980   &   4.4   &   5.81   &    \\
             &          &        &           &        &      &     \\
Cu~{\sc i}    &  $ N=2$   &      &      &      &   3.56$\pm 0.05$   &   \\
 \hline 
29.00   &   5105.537   &   1.389   &   $-$1.516   &   3.8   &   3.51   &    \\
29.00   &   5218.197   &   3.817   &   0.476   &   7.0   &   3.60   &    \\
   &      &      &      &      &      &    \\
Zn~{\sc i}     &  $ N=3$   &      &      &      &   4.15$\pm 0.03$   &   \\
 \hline 
30.00   &   4680.134   &   4.006   &   $-$0.815   &   8.1   &   4.11   &    \\
30.00   &   4722.153   &   4.030   &   $-$0.338   &   23.7   &   4.17   &    \\
30.00   &   4810.528   &   4.078   &   $-$0.137   &   32.2   &   4.18   &    \\
   &      &      &      &      &      &    \\
Y~{\sc ii}    &  $ N=7$   &      &      &      &   1.87$\pm 0.06$   &   \\
 \hline 
39.01   &   4883.684   &   1.084   &   0.070   &   39.6   &   1.88   &    \\
39.01   &   5087.416   &   1.084   &   $-$0.170   &   29.8   &   1.90   &    \\
39.01   &   5123.211   &   0.992   &   $-$0.830   &   7.9   &   1.84   &    \\
39.01   &   5200.406   &   0.992   &   $-$0.570   &   12.6   &   1.79   &    \\
39.01   &   5402.774   &   1.839   &   $-$0.510   &   4.3   &   1.88   &    \\
39.01   &   5509.895   &   0.992   &   $-$1.010   &   5.1   &   1.80   &    \\
39.01   &   5662.925   &   1.944   &   0.160   &   19.2   &   1.98   &    \\
   &      &      &      &      &      &    \\
Zr~{\sc ii}    &  $ N=1$   &      &      &      &   2.38   &   \\
 \hline 
40.01   &   5112.297   &   1.665   &   $-$0.590   &   5.0   &   2.38   &    \\
   &      &      &      &      &      &    \\
Ba~{\sc ii}   &  $ N=4$   &      &      &      &   1.58$\pm 0.04$   &   \\
 \hline 
56.01   &   4553.995   &   0.000   &   0.163   &   131.1   &   1.52   &
 hfs fit \\
56.01   &   4934.076   &   0.000   &   $-$0.150   &   113.8   &   1.58   &
 hfs fit \\ 
56.01   &   5853.668   &   0.604   &   $-$1.000   &   21.6   &   1.62   &
 hfs fit \\
56.01   &   6141.713   &   0.704   &   $-$0.076   &   70.8   &   1.60   &
 hfs fit \\
   &      &      &      &      &      &    \\
Ce~{\sc ii}   &  $ N=1 $  &      &      &      &   1.26   &    \\  
 \hline
58.01   &   4628.161   &   0.516   &   0.008   &   2.2   &   1.26   &    \\
   &      &      &      &      &      &    \\
Nd~{\sc ii}   &  $ N=1$   &      &      &      &   0.91   &     \\
 \hline 
60.01   &   5130.586   &   1.304   &   0.570   &   1.4   &   0.91   &     \\

\bottomrule
\end{tabular}
\label{tab:A4}
\end{table*}

\end{document}